\documentclass[
 reprint,
 superscriptaddress,
 amsmath,amssymb,
 aps,
prb,
]{revtex4-2}

\usepackage{graphicx}
\usepackage{subcaption}
\usepackage{dcolumn}
\usepackage{bm}
\usepackage{pgf}
\usepackage{physics}
\usepackage{comment}
\usepackage[format=plain,justification=justified]{caption}
\usepackage{dsfont}
\usepackage{xcolor}
\usepackage{float}
\usepackage{ragged2e}
\usepackage{bbold}
\usepackage{amsfonts}
\usepackage{hyperref}
    \hypersetup{colorlinks, citecolor=black, linkcolor= black, urlcolor=blue}

\newcommand\numberthis{\addtocounter{equation}{1}\tag{\theequation}}

\begin{document}
\title{Boltzmann-Bloch Equation Approach to the Theory of the Optical Inter- and Intraband Response in Noble Metals}
\author{Robert Lemke}
\email{r.lemke@campus.tu-berlin.de}
\affiliation{Nichtlineare Optik und Quantenelektronik, Institut für Physik und Astronomie, Technische Universität Berlin, 10623 Berlin, Germany}
\author{Matthias Rössle}
\affiliation{Helmholtz-Zentrum Berlin für Materialien und Energie, 14109 Berlin, Germany}
\author{Holger Lange}
\affiliation{Institut f\"ur Physik und Astronomie, Universit\"at Potsdam, 14476 Potsdam, Germany}
\author{Andreas Knorr}
\affiliation{Nichtlineare Optik und Quantenelektronik, Institut für Physik und Astronomie, Technische Universität Berlin, 10623 Berlin, Germany}
\author{Jonas Grumm}
\affiliation{Nichtlineare Optik und Quantenelektronik, Institut für Physik und Astronomie, Technische Universität Berlin, 10623 Berlin, Germany}
\affiliation{Institut f\"ur Physik und Astronomie, Universit\"at Potsdam, 14476 Potsdam, Germany}
\date{\today}
\begin{abstract}

In this paper we introduce momentum-resolved metal Boltzmann-Bloch equations (MBBE) for the combined description of electronic intra- and interband processes in noble metals.
This microscopic framework incorporates a full treatment of many-body electron-electron and electron-phonon interactions, relevant for relaxation and dephasing processes after optical excitation.
For the example of gold, we calculate the linear optical response for near-infrared and visible energies. This provides insight into the interplay of microscopic processes hidden in phenomenological Drude-Lorentz models. 
The complex geometry of the Fermi surface is treated by an anisotropic electronic dispersion model, which is necessary to explain the temperature dependent spectrum over the whole frequency range of intra- and interband transitions.
\\

\end{abstract}
\maketitle

\section{\label{sec:intro} Introduction}

\begin{figure}[ht]
    \centering
    \includegraphics[width=\linewidth]{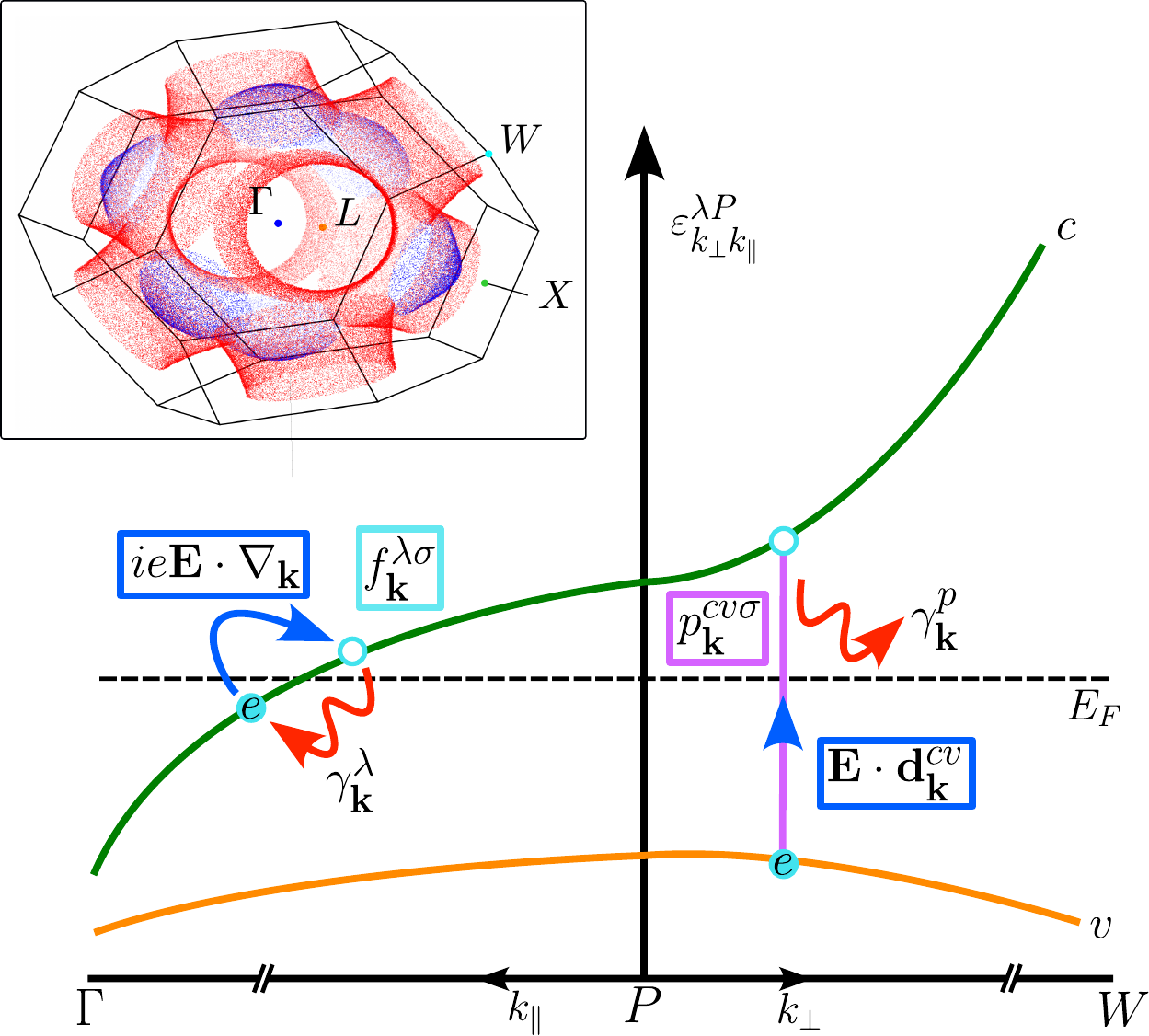}
    \caption{\justifying Schematic illustration of the anisotropic dispersion relation $\varepsilon^{\lambda P}_{k_{\perp}k_{\parallel}}$ in the vicinity of high symmetry points $P = X,L$ of the valence band (orange, $\lambda = v$) and conduction band (green, $\lambda = c$) with Fermi energy $E_F$. The light blue circles and the purple line respectively depict the electron occupations $f_{\mathbf{k}}^{\lambda\sigma}$ and the interband transitions $p_{\mathbf{k}}^{vc\sigma}$. 
    The blue arrows account for excitation processes from occupied states into unoccupied states either within the conduction band (intraband processes) or from the valence into the conduction band (interband processes) induced by an external electrical field $\mathbf{E}$, cp.~discussion in Sec.~\ref{sec:theo_background}. Many-body relaxation ($\gamma_{\mathbf{k}}^{\lambda}$) of the occupations and dephasing processes ($\gamma_{\mathbf{k}}^{p}$) of the interband transitions by electron-phonon and electron-electron scattering processes, are schematically indicated by the red arrows. The Fermi surface of the anisotropic dispersion model is shown in the upper left corner, assembled from the contributions of all fourteen $X$ and $L$ sections of the fcc Brillouin zone.}
    \label{fig:theo:dispersion_skizze}
\end{figure}

Plasmonic applications leverage the light-matter interaction in noble metal nano-structures with relevance ranging from molecular sensing, solar energy harvesting, and plasmonic modulators to light-driven chemical catalysis  \cite{bigdeli_nanoparticle-based_2017,jang_plasmonic_2016,dong_plasmonic_2023,de_aberasturi_modern_2015,tokel_advances_2014,hartland_plasmonics_2025,jain_noble_2008,klein_2d_2019,herran_plasmonic_2023}. 
Since the response of plasmonic nanostructures is based on understanding the bulk properties of the metal, accompanied by boundary conditions for the Maxwell equations, microscopic knowledge of the dielectric function $\epsilon(\omega)$ is of great importance. Additionally, from a theoretical point of view, the description of plasmonic effects naturally focuses on the study of conduction band electrons, referred to as intraband processes. Interband excitations from valence bands into the conduction band are often addressed by phenomenological approaches to the dielectric function \cite{etchegoin_analytic_2006,pfeifer_time-domain_2024}, hiding the underlying microscopical nature of these mechanisms, making a synergistic view  of both types of processes unfeasible \cite{fu_bridging_2026}. 
A general microscopic framework starting from a wave number resolved Bloch-electron formulation, which focuses on the interplay of both intra- and interband excitation, as well as relaxation and dephasing processes in noble metals is, to our knowledge, missing. This work is built on Boltzmann equation approaches for electron occupations in the conduction band \cite{rethfeld_ultrafast_2002,del_fatti_nonequilibrium_2000} and extends it to interband transitions. \\
Historically, one of the first classical descriptions of the electrical properties of metals was provided by Drude \cite{drude_zur_1900}, assuming a free electron gas with mean collision time in an electrical field. In the frame of quantum mechanics, an analogy can be drawn to the acceleration and scattering of conduction band electrons, schematically illustrated by the blue and red arrow on the left hand side of Fig.~\ref{fig:theo:dispersion_skizze} respectively. \\
The role of interband transitions, which do not have a classical correspondence (purple line in Fig.~\ref{fig:theo:dispersion_skizze}), was identified in 1965 by Cooper et al. \cite{cooper_optical_1965} in their analysis of the rise in the absorption of noble metals in the low energetic visible spectrum.
For semiconductors, Lindberg and Koch developed the semiconductor Bloch equations (SBE) \cite{lindberg_effective_1988}, using quantum mechanical projector-operator techniques, taking into account many-body effects. This provided a theoretical cornerstone framework for the modern description of optical properties of semiconductor structures.
In contrast to semiconductors, noble metals feature a partially filled conduction band, which means that a perturbation of the electronic ground state carries the signature of complex geometrical features of the Fermi surface.

In this work, we apply the microscopic methods of the SBE to noble metals and introduce metal Boltzmann-Bloch equations (MBBE), which establish a microscopic framework for the electron-electron and electron-phonon scattering dynamics.\\
As a first example of application, we apply the MBBE framework to study the linear optical response of bulk gold. Therefore, the geometry of the Fermi surface is addressed by employing an anisotropic electronic dispersion model \cite{cooper_optical_1965,guerrisi_splitting_1975,rosei_d_1973}, cp. Fig.~\ref{fig:theo:dispersion_skizze}.\\
This article is structured as follows: we start with the theoretical framework in Sec.~\ref{sec:theo_background} and derive the metal Boltzmann-Bloch equations including electron-phonon and electron-electron many-body scattering from a general noble metal Hamiltonian. In Sec.~\ref{sec:dispersion_relation}, we introduce an anisotropic electronic band structure model to address the momentum-dependence of the conduction- and valence band energy realistically. In Sec.~\ref{sec:linearization} we linearize the MBBE for the calculation of spectra and present semi-analytical expressions for electron-phonon and electron-electron relaxation and dephasing rates. Lastly, in Sec.~\ref{sec:results}, we discuss the signatures of the anisotropic model and discuss temperature dependent semi-analytical relaxation and dephasing rates as well as the related linear spectra. The latter are compared with experimental data measured in our own ellipsometry experiments.

\section{\label{sec:theo_background} Theoretical Framework}
The aim of this section is to introduce the formalism for the description of the coupled dynamics of electronic occupations and interband transitions in a momentum resolved two band model of noble metals after optical excitation, cp. Fig.~\ref{fig:theo:dispersion_skizze}. To address relaxation and dephasing phenomena, we include electron-electron and electron-phonon interactions in a many-body framework.

\subsection{Microscopic Approach}\label{sec:theo_background:micro}
Starting point for the description of electronic excitations in the valence ($v$) and conduction ($c$) band in noble metals is a two-band Hamiltonian, \cite{meier_coherent_1994, kira_quantum_1999, rossi_theory_2002, axt_femtosecond_2004, kira_semiconductor_2011}
cp. Fig.~\ref{fig:theo:dispersion_skizze},
\begin{align}
    &H = \sum_{\lambda\mathbf{k}\sigma} \varepsilon_{\mathbf{k}}^{\lambda} a_{\mathbf{k}}^{\lambda\sigma\dagger} a_{\mathbf{k}}^{\lambda\sigma} \label{eq:hamiltonian}\\
    +&ie\sum_{\lambda\mathbf{kq}\sigma} \mathbf{E}_\mathbf{\text{-}q} \cdot a_{\mathbf{k}}^{\lambda\sigma\dagger} \nabla_\mathbf{k} [a_{\mathbf{k+q}}^{\lambda\sigma}]\nonumber - \sum_{\lambda\lambda' \mathbf{k q}\sigma}\mathbf{E}_{\text{-}\mathbf{q}}\cdot\mathbf{d}^{\lambda\lambda'}_{\mathbf{k}}a_{\mathbf{k}}^{\lambda\sigma\dagger}a_{\mathbf{k+q}}^{\lambda'\sigma}\\  
    + & \sum_{\mathbf{q}} \hbar \omega_{\mathbf{q}} b^\dagger_{\mathbf{q} } b_{\mathbf{q} }+ \sum_{\substack{\lambda\sigma\\\mathbf{kqG}}} g_{\mathbf{qG}}^{\lambda} (b^\dagger_{\mathbf{-q}} + b_{\mathbf{q}}) a_{\mathbf{k+q+G}}^{\lambda\sigma\dagger}a_{\mathbf{k}}^{\lambda\sigma} - H_{W}^{ep}  \nonumber\\
    + &\sum_{\substack{\lambda\lambda'\mathbf{qkk}'\\ \sigma\sigma'\mathbf{GG'}}}V_{\mathbf{q}}^{\lambda\lambda'}a_{\mathbf{k+q+G}}^{\lambda\sigma\dagger}a_{\mathbf{k'-q-G'}}^{\lambda'\sigma'\dagger}a_{\mathbf{k'}}^{\lambda'\sigma'}a_{\mathbf{k}}^{\lambda\sigma} - H_{MF}^{ee} ~\nonumber.
\end{align}
The first line in Eq.~(\ref{eq:hamiltonian}) represents the single-particle electron Hamiltonian with spin degenerate dispersion $\varepsilon_{\mathbf{k}}^{\lambda}$ and band index $\lambda \in \{v,c\}$, which will be discussed in Sec.~\ref{sec:dispersion_relation}. Here, $a_{\mathbf{k}}^{\lambda\sigma(\dagger)}$ designate fermionic electron annihilation (creation) operators with lattice momentum $\mathbf{k}$, band $\lambda$ and spin $\sigma$. 
The two contributions in the second line correspond to the semi-classical light-matter intra- and interband interaction with the Fourier component of the optical field $\mathbf{E}_{\text{-}\mathbf{q}}$. Here, $-ie \nabla_\mathbf{k}$ describes the intraband dipole operator \cite{adams_crystal_1953} and $\mathbf{d}_{\mathbf{k}}^{\lambda\lambda'}$ is the interband dipole matrix element \cite{haug_quantum_2008}. The first term in the third line accounts for the dispersion of longitudinal acoustic (LA) phonons with bosonic annihilation (creation) operators $b_{\mathbf{q}}^{(\dagger)}$, momentum $\mathbf{q}$ and phonon dispersion $\omega_{\mathbf{q}} = c_{LA}|\mathbf{q}|$  in Debye approximation \cite{czycholl_basics_2023}, where $c_{LA}$ denotes the velocity of sound. The second term designates the electron-phonon interaction and includes Umklapp processes specified by the reciprocal lattice vectors $\mathbf{G}$ \cite{ashcroft_solid_1976,czycholl_basics_2023}.  This interaction describes the transfer of momentum and energy from the ion lattice to an electron or vice versa by either emission or absorption of a phonon with momentum $\vb{q}$. We apply a screened matrix element for electron-phonon interaction \cite{ashcroft_solid_1976}
\begin{align}
    |g_{\mathbf{qG}}^{\lambda}|^2 = \frac{e^2}  {V2\varepsilon_0}\frac{\hbar\omega_{\mathbf{q}}}{|\mathbf{q+G}|^2 + \kappa_{TF}^2} \label{eq:theory:phonon_coupling}, 
\end{align} 
where $\kappa_{TF}$ designates the Thomas-Fermi wave number \cite{ashcroft_solid_1976},

\begin{align}
   \kappa_{TF}^2 = \frac{e^2}{\varepsilon_0}\partial_{\mu}n_0~,
\end{align}
with $n_0$ the equilibrium electron density of the conduction band and $\mu$ the chemical potential. The contribution of transversal acoustic (TA) phonons is omitted as the coupling strength is negligible compared to LA phonons \cite{ashcroft_solid_1976}. The third term $H_W^{ep}$ is introduced to ensure that the interacting many-body system of phonons and electrons in the filled valence and partially filled conduction band behaves stationary in the ground state \cite{weber_optical_2007}. 
For further details we refer to App.~\ref{app:weyl}. 
\\
The first contribution in the last line corresponds to electron-electron interaction with the screened Coulomb matrix element \cite{czycholl_basics_2023,ashcroft_solid_1976}
\begin{align}
    V_{\mathbf{q}}^{\lambda\lambda'} = \frac{e^2}{V\varepsilon_0}\frac{1}{|\mathbf{q}|^2 + \kappa_{TF}^2}~ \label{eq:theory:electron_coupling}.
\end{align}
This interaction conserves momentum between scattered electrons up to two independent reciprocal lattice vectors $\mathbf{G}$ and $\vb{G}'$. In this approach, we apply matrix elements in Eqs.~(\ref{eq:theory:phonon_coupling},\ref{eq:theory:electron_coupling}) as independent from the band.
\\ 
The second term $H_{MF}^{ee}$ subtracts the electron mean-field in the equilibrium, which is already considered in the applied electron dispersion $\varepsilon^\lambda_\mathbf{k}$ \cite{rangel_band_2012}, cf.~Eq.~\eqref{eq:theo:mean_field_polarizations}.
\\
The macroscopic polarization $\mathbf{P}$, retrieved from the Hamiltonian in Eq.~(\ref{eq:hamiltonian}) by
\begin{align}
    \mathbf{P}(t) = \mathbf{P}_{\text{intra}}(t) + \mathbf{P}_{\text{inter}}(t)=  -\langle \delta H(t)/\delta \mathbf{E}(t)\rangle ~, 
\end{align}
 is connected with the microscopic electron occupations $f_{\mathbf{k}}^{\lambda \sigma} =  \langle a_{\mathbf{k}}^{\lambda\sigma\dagger}a_{\mathbf{k}}^{\lambda\sigma}\rangle$ and interband transitions $p_{\mathbf{k}}^{vc\sigma} = \langle a_{\mathbf{k}}^{ v \sigma \dagger}a_{\mathbf{k}}^{ c \sigma}\rangle$.
For spatially homogeneous excitations, i.e in local response approximation for the optical fields $\mathbf{E}_{\text{-}\mathbf{q}}(t) = \delta_{\mathbf{q,0}}\mathbf{E}(t)$, the intraband polarization is characterized by the current density
\begin{align}
    \mathbf{j}(t) = \partial_t \mathbf{P}_{\text{intra}}(t) = -\frac{e}{V}\sum_{\lambda\sigma\mathbf{k}}\mathbf{v}_{\mathbf{k}}^{\lambda}f_{\mathbf{k}}^{\lambda\sigma}(t) \label{eq:theo:p_intra},
\end{align}
with the electron velocity $\mathbf{v}_{\mathbf{k}}^{\lambda} = \frac{1}{\hbar}\nabla_{\mathbf{k}}\varepsilon_{\mathbf{k}}^{\lambda}$. Similarly, the interband polarization reads
\begin{align}
    \mathbf{P}_{\text{inter}}(t) = \frac{1}{V}\sum_{\sigma\mathbf{k}}\mathbf{d}_{\mathbf{k}}^{vc}p_{\mathbf{k}}^{ vc\sigma}(t) + c.c. \label{eq:theo:p_inter}~.
\end{align}
In the next section, we derive the set of coupled equations of motion for $p_{\mathbf{k}}^{vc\sigma}$ and $f_{\mathbf{k}}^{\lambda\sigma}$ representing the metal Boltzmann-Bloch equations. They contain light-matter interactions (Sec. \ref{sec:theo:light_matter}) and many-body scattering, (Sec. \ref{sec:theo_scattering}), derived in the Heisenberg picture.
\subsection{Light-Matter Interaction}\label{sec:theo:light_matter}

The excitation dynamics of the electron occupation in each band, due to the light-matter interaction ($lm$), stems from the first two lines of the Hamiltonian in Eq.~\eqref{eq:hamiltonian},
\begin{align}
     \dot{f}_{\mathbf{k}}^{\lambda\sigma}(t)\big\vert_{lm} = &\mathbf{E}(t)\cdot \Big[ \frac{e}{\hbar}\nabla_{\mathbf{k}}f_{\mathbf{k}}^{\lambda\sigma}(t) + \frac{2}{\hbar}\Im\{\mathbf{d}_{\mathbf{k}}^{vc}p_{\mathbf{k}}^{ vc\sigma}(t) \}\Big] ~. \label{eq:theo_background:boltzmann}
\end{align}
The first term in Eq.~(\ref{eq:theo_background:boltzmann}) arises from the intraband light-matter interaction, resulting in an acceleration of electrons in the polarization direction of the optical field \cite{grumm_femtosecond_2025,grumm_theory_2025}. In the limit of linear optics, this contribution leads to the well-known Drude response of the quasi-free conduction electrons. The second term describes an excitation of electron occupations by the dipole coupling between the optical field and interband transitions.
\\ 
The corresponding dynamics of the interband transitions read
\begin{align*}
    \dot{p}^{vc\sigma}_{\mathbf{k}}(t)\big\vert_{lm} =& \frac{i}{\hbar}(\varepsilon_{\mathbf{k}}^{v} - \varepsilon_{\mathbf{k}}^{c})p_{\mathbf{k}}^{ vc\sigma}(t) \label{eq:theo_background:bloch} \numberthis \\
    +&\mathbf{E}(t)\cdot \Big[ \frac{e}{\hbar}\nabla_{\mathbf{k}}p_{\mathbf{k}}^{ vc\sigma}(t) + \frac{i}{\hbar}\mathbf{d}^{cv}_{\mathbf{k}}(f_{\mathbf{k}}^{ v\sigma}(t) - f_{\mathbf{k}}^{ c\sigma}(t)) \Big]~. 
\end{align*}
The first term in Eq.~(\ref{eq:theo_background:bloch}) constitutes the single-particle transition energies for excitations at and above the energy band gap $\varepsilon_{\mathbf{k}}^c - \varepsilon_{\mathbf{k}}^v $. The second term refers to a momentum shift of an optically generated interband transition in momentum space and is therefore an intrinsic non-linear process. Lastly, interband transitions are induced directly by the third term, for non-zero electron occupation differences between valence and conduction band \cite{haug_quantum_2008, lindberg_effective_1988}. This term already contributes for linear optics and depends due to the partially filled conduction band on the shape of the Fermi surface. This is in contrast to semiconductors with empty conduction band and Fermi level in the band gap.
\subsection{Many-Body Interactions}\label{sec:theo_scattering}
To address energy and momentum relaxation of the electronic occupations and dephasing of interband transitions, via many-body interactions during and after optical excitation, we account for electron-phonon and electron-electron interaction. 
Their contribution to the MBBE is derived by evaluating the Heisenberg equations of motion with respect to the third and fourth lines of  the Hamiltonian, Eq.~\eqref{eq:hamiltonian}. 
The hierarchy problems arising here are truncated within a correlation expansion and a Markov approximation \cite{grumm_theory_2025, haug_quantum_2008, fricke_transport_1996}. Electron-phonon mean-field contributions lead to a renormalization of the interband transition resonance $(\varepsilon_{\mathbf{k}}^{v} - \varepsilon_{\mathbf{k}}^{c})/\hbar$ and will be neglected here.\\
In the case of electron-electron interaction, coherent mean field contributions arise, resulting from a Hartree-Fock factorization ($\lambda' \neq \lambda$), i.e.
\begin{align}
    \dot{f}_{\mathbf{k}}^{\lambda\sigma}(t)\vert_{HF} =  &\frac{i}{\hbar}\sum_{\mathbf{qG}} V_{\vb{q}}^{\lambda'\lambda}p_{\vb{k+q+G}}^{\lambda'\lambda}(t)p_{\vb{k}}^{\lambda\lambda'}(t) - c.c.~,\label{eq:theo:mean_field_occupations}\\
    \dot{p}_{\mathbf{k}}^{vc\sigma}(t)\vert_{HF} = &\frac{i}{\hbar}\sum_{\vb{qG}}(V_{\vb{q}}^{vc}f_{\vb{k}}^{v\sigma}(t) - V_{\vb{q}}^{cv}f_{\vb{k}}^{c\sigma}(t))p_{\vb{k+q+G}}^{vc\sigma}(t) \nonumber\\
     &+\frac{i}{\hbar}\sum_{\vb{qG}}\big(V_{\vb{q}}^{cc}(f_{\vb{k+q+G}}^{c\sigma}(t) - f_{\vb{k+q+G}}^{c\sigma,eq}) \label{eq:theo:mean_field_polarizations}\\
     &\qquad - V_{\vb{q}}^{vv}(f_{\vb{k+q+G}}^{v\sigma}(t) - f_{\vb{k+q+G}}^{v\sigma,eq})\big)p_{\vb{k}}^{vc\sigma}(t) \nonumber~.
\end{align} 
We note that the second and third lines of Eq.~(\ref{eq:theo:mean_field_polarizations}) involve only the optically induced non-equilibrium electron occupations, as equilibrium contributions, incorporated by $H_{MF}^{ee}$ \cite{hess_maxwell-bloch_1996}, are already considered in the band structure \cite{rangel_band_2012}.\\
Within a Born-Markov approximation \cite{rossi_theory_2002, haug_quantum_2008}, the second order correlations yield the scattering contributions ($sc$) and take the form of Boltzmann collision terms
\begin{align}
    \dot{f}_{\mathbf{k}}^{\lambda\sigma}\vert_{sc} &= -\Gamma_{\mathbf{k}}^{\lambda\sigma,out}f_{\mathbf{k}}^{\lambda\sigma} + \Gamma_{\mathbf{k}}^{\lambda\sigma,in}(1-f_{\mathbf{k}}^{\lambda\sigma}) + \Gamma_{\mathbf{k}}^{\lambda\sigma,nl} ~,\label{eq:theo_scattering_occupation_scattering} \\
    \dot{p}_{\mathbf{k}}^{ vc\sigma}\vert_{sc} &= -\Gamma_{\mathbf{k}}^{p\sigma,d} p_{\mathbf{k}}^{ vc\sigma} + \sum_{\mathbf{qG}}\Gamma_{\mathbf{kq}}^{p\sigma,nd} p_{\mathbf{k+q+G}}^{vc\sigma}  \nonumber \\
    &\hspace{3.5cm}+ T_{\mathbf{k}}^{p\sigma,ee} + \Gamma_{\mathbf{k}}^{p\sigma,nl} ~, \label{eq:theo_scattering_transition_scattering}
\end{align}
where each rate is composed of electron-phonon ($ep$) and electron-electron ($ee$) contributions, i.e $\Gamma_\mathbf{k} = \Gamma^{ep}_\mathbf{k} + \Gamma^{ee}_\mathbf{k}$.
\\
For electron occupations $f_{\mathbf{k}}^{\lambda\sigma}$ the electron-electron scattering rates in Eq.~\eqref{eq:theo_scattering_occupation_scattering} read 
\begin{align*}
    &\Gamma_{\mathbf{k}}^{\lambda\sigma,out,ee} = \numberthis \label{eq:theo_scattering_occupation_ee_out} \\
    &\frac{2\pi}{\hbar}\sum_{\substack{\mathbf{qk'GG'}\\\sigma'\lambda'}}(V_{\mathbf{q}}^{\lambda'\lambda} - \delta_{\lambda\lambda'}^{\sigma\sigma'}V_{\mathbf{k'-q-k}}^{\lambda\lambda} )V_{\mathbf{q}}^{\lambda\lambda'}f_{\mathbf{k'}}^{\lambda'\sigma'}(1-f_{\mathbf{k'-q-G'}}^{\lambda'\sigma'})\times
    \\&\times(1-f_{\mathbf{k+q+G}}^{\lambda\sigma})\delta(\varepsilon_{\mathbf{k}}^{\lambda} + \varepsilon_{\mathbf{k'}}^{\lambda'} - \varepsilon_{\mathbf{k+q+G}}^{\lambda} - \varepsilon_{\mathbf{k'-q-G'}}^{\lambda'})
     ~, 
\end{align*}
\begin{align*}
    &\Gamma_{\mathbf{k}}^{\lambda\sigma,in,ee} = \numberthis \label{eq:theo_scattering_occupation_ee_in}  \\
    &\frac{2\pi}{\hbar}\sum_{\substack{\mathbf{qk'GG'}\\\sigma'\lambda'}}(V_{\mathbf{q}}^{\lambda'\lambda} - \delta_{\lambda\lambda'}^{\sigma\sigma'}V_{\mathbf{k'-q-k}}^{\lambda\lambda})V_{\mathbf{q}}^{\lambda\lambda'}(1-f_{\mathbf{k'}}^{\lambda'\sigma'})f_{\mathbf{k'-q-G'}}^{\lambda'\sigma'}\times
    \\&\times f_{\mathbf{k+q+G}}^{\lambda\sigma}\delta(\varepsilon_{\mathbf{k}}^{\lambda} + \varepsilon_{\mathbf{k'}}^{\lambda'} - \varepsilon_{\mathbf{k+q+G}}^{\lambda} - \varepsilon_{\mathbf{k'-q-G'}}^{\lambda'}) ~. 
\end{align*}
Here, electrons in bands $\lambda$ and $\lambda'$ with momenta $\mathbf{k}$ and $\mathbf{k'}$ scatter within their bands into states $\mathbf{k+q+G}$ and $\mathbf{k'-q-G'}$, respectively, under the exchange of momentum $\mathbf{q}$ and conservation of energy. The reciprocal lattice vectors $\mathbf{G}$ and $\mathbf{G'}$ allow a restriction of all momenta to the first Brillouin zone, whereby the total momentum is only quasi-conserved. With the correlation expansion, the coupling arises from a direct contribution ($V_{\mathbf{q}}^{\lambda\lambda'}$), as well as an exchange contribution ($V_{\mathbf{k'-q-k}}^{\lambda\lambda}$), following from Pauli repulsion for interacting electrons with same band and spin ($\sim \delta_{\lambda\lambda'}^{\sigma\sigma'}$), reducing the scattering amplitude.
\\
The electron-phonon scattering rates in Eq.~(\ref{eq:theo_scattering_occupation_scattering}) read
\begin{align}
    \Gamma_{\mathbf{k}}^{\lambda\sigma,in,ep} = \frac{2\pi}{\hbar}\sum_{\mathbf{qG\pm}}\Big(n_{\mathbf{q}} + \frac{1}{2} \pm \frac{1}{2}\Big)|g_\mathbf{qG}^{\lambda}|^2 \times \nonumber\\
    \times\delta(\varepsilon_{\mathbf{k+q+G}}^\lambda - \varepsilon_{\mathbf{k}}^\lambda \pm \hbar \omega_{\mathbf{q}})f_{\mathbf{k+q+G}}^{\lambda\sigma}, \label{eq:theo_scattering_occupation_ep_out}
\end{align}
\begin{align}
    \Gamma_{\mathbf{k}}^{\lambda\sigma,out,ep} = \frac{2\pi}{\hbar}\sum_{\mathbf{qG\pm}}\Big(n_{\mathbf{q}} + \frac{1}{2} \mp \frac{1}{2}\Big)|g_\mathbf{qG}^{\lambda}|^2\times \nonumber\\
    \times\delta(\varepsilon_{\mathbf{k+q+G}}^{\lambda} - \varepsilon_{\mathbf{k}}^{\lambda} \pm \hbar \omega_{\mathbf{q}})(1-f_{\mathbf{k+q+G}}^{\lambda\sigma}), \label{eq:theo_scattering_occupation_ep_in}
\end{align}
where $n_{\mathbf{q}} = \langle b_{\mathbf{q}}^{\dagger}b_{\mathbf{q}}\rangle $ is the phonon occupations. Here, an electron with momentum $\mathbf{k}$ scatters within the same band $\lambda$ into the state $\mathbf{k+q+G}$ (or vice versa) by either absorbing or emitting a phonon with momentum $\mathbf{q}$, again under conservation of energy and momentum up to a reciprocal lattice vector $\mathbf{G}$. Phonon-assisted interband transitions $\lambda\rightarrow \lambda'$ do not occur as they are energetically improbable in noble metals with band gap energies in the optical range.
\\
For the phonon occupations applies the equation of motion \cite{richter_two-dimensional_2009}
\begin{align*}
    &\dot{n}_{\mathbf{q}} = \frac{2\pi}{\hbar}\sum_{\substack{\lambda,\lambda'\neq \lambda\\\sigma\mathbf{kG}}}\bigg\{|g_{\mathbf{qG}}^{\lambda}|^2 \{(n_\mathbf{q} + 1)f_{\mathbf{k+q + G}}^{\lambda\sigma}(1-f_{\mathbf{k}}^{\lambda\sigma}) \\
    &- n_{\mathbf{q}} f_{\mathbf{k}}^{\lambda\sigma}(1-f_{\mathbf{k+q+G}}^{\lambda\sigma})\} - \text{Re}(g_{\mathbf{qG}}^{\lambda'}g_{\mathbf{qG}}^{\lambda} p_{\mathbf{k}}^{\lambda'\lambda\sigma}p_{\mathbf{k+q+G}}^{\lambda\lambda'\sigma})\bigg\}\times\\
    &\qquad\qquad \times\delta(\varepsilon_{\mathbf{k+q+G}}^{\lambda} - \varepsilon_{\mathbf{k}}^{\lambda} - \hbar\omega_{\mathbf{q}}).\numberthis\label{eq:eom_phonon_occupations}
\end{align*}
Here, the first two terms reflect emission or absorption of a phonon by the electronic system under conservation of energy. The last term accounts for interband transition effects. The exchange of thermal energy between electrons and phonons can be approximated for example by a two-temperature model after thermalization \cite{allen_theory_1987,rethfeld_ultrafast_2002}.
For the scope this work, focusing on linear optics, energy transfer to the phonons is negligible, i.e.~we will set $\dot{n}_\mathbf{q}=0$ within a bath approximation in Sec.~\ref{sec:linearization}.

The interband transition electron-electron scattering rates in Eq.~\eqref{eq:theo_scattering_transition_scattering} read 
\begin{align}
    &\Gamma_{\mathbf{k}}^{p\sigma,d,ee} = \frac{\pi}{\hbar}\sum_{\substack{\mathbf{qk'GG'}\\\sigma'\lambda\lambda'}}(V_{\mathbf{q}}^{\lambda'\lambda} - \delta_{\lambda\lambda'}^{\sigma\sigma'}V_{\mathbf{k'-q-k}}^{\lambda\lambda})V_{\mathbf{q}}^{\lambda\lambda'}\times \label{eq:theo_scattering_transition_ee_d}\\
    &\times \{(1-f_{\mathbf{k'}}^{\lambda'\sigma'}) f_{\mathbf{k'-q-G'}}^{\lambda'\sigma'}f_{\mathbf{k+q+G}}^{\lambda\sigma} + f_{\mathbf{k'}}^{\lambda'\sigma'}(1-f_{\mathbf{k'-q-G'}}^{\lambda'\sigma'})\times\nonumber\\
    &\times(1-f_{\mathbf{k+q+G}}^{\lambda\sigma})\} \delta(\varepsilon_{\mathbf{k}}^{\lambda} + \varepsilon_{\mathbf{k'}}^{\lambda'} - \varepsilon_{\mathbf{k+q+G}}^{\lambda} - \varepsilon_{\mathbf{k'-q-G'}}^{\lambda'})~,\nonumber
\end{align}
\begin{align}
    &\Gamma_{\mathbf{kq}}^{p\sigma,nd,ee}  = \frac{\pi}{\hbar}\sum_{\substack{\mathbf{k'GG'}\\\sigma'\lambda\lambda'}}(V_{\mathbf{q}}^{\lambda'\lambda} - \delta_{\lambda\lambda'}^{\sigma\sigma'}V_{\mathbf{k'-q-k}}^{\lambda\lambda})V_{\mathbf{q}}^{\lambda\lambda'}\times \label{eq:theo_scattering_transition_ee_nd}\\
    &\times\{(1-f_{\mathbf{k'}}^{\lambda'\sigma'})f_{\mathbf{k'-q-G'}}^{\lambda'\sigma'}(1-f_{\mathbf{k}}^{\lambda\sigma}) + f_{\mathbf{k'}}^{\lambda'\sigma'}(1-f_{\mathbf{k'-q-G'}}^{\lambda'\sigma'})\times\nonumber\\ 
    &\times f_{\mathbf{k}}^{\lambda\sigma} \}
    \delta(\varepsilon_{\mathbf{k}}^{\lambda} + \varepsilon_{\mathbf{k'}}^{\lambda'} - \varepsilon_{\mathbf{k+q+G}}^{\lambda} - \varepsilon_{\mathbf{k'-q-G'}}^{\lambda'})~, \nonumber
\end{align}
and include direct and exchange contributions as for the electron occupations.\\
The interband transition electron-phonon scattering rates in Eq.~\eqref{eq:theo_scattering_transition_scattering} read
\begin{align}
    \Gamma_{\mathbf{k}}^{p\sigma,d,ep} =& \frac{\pi}{\hbar}\sum_{\substack{\mathbf{qG}\\\lambda\pm}}|g_{\mathbf{qG}}^{\lambda}|^2\{(1+n_{\mathbf{q}})\Big(\frac{1}{2} \pm \frac{1}{2} \mp f_{\mathbf{k+q+G}}^{\lambda\sigma}\Big)\nonumber \\
    +& n_{\mathbf{q}}\Big(\frac{1}{2} \mp \frac{1}{2} \pm f_{\mathbf{k+q+G}}^{\lambda\sigma}\Big)\}\delta(\varepsilon_{\mathbf{k+q+G}}^{\lambda} - \varepsilon_{\mathbf{k}}^{\lambda} \pm \hbar\omega_{\mathbf{q}})~,\label{eq:theo_scattering_transition_ep_d}
\end{align}
\begin{align}
    \Gamma_{\mathbf{kq}}^{p\sigma,nd,ep} =& \frac{\pi}{\hbar}\sum_{\substack{\mathbf{G}\lambda\pm}}|g_{\mathbf{qG}}^{\lambda}g_{\mathbf{qG}}^{\lambda'}|\{(1+n_{\mathbf{q}})\Big(\frac{1}{2} \mp \frac{1}{2} \pm f_{\mathbf{k}}^{\lambda\sigma}\Big)\nonumber \\
    +& n_{\mathbf{q}}\Big(\frac{1}{2} \pm \frac{1}{2} \mp f_{\mathbf{k}}^{\lambda\sigma}\Big)\}\delta(\varepsilon_{\mathbf{k+q+G}}^{\lambda} - \varepsilon_{\mathbf{k}}^{\lambda} \pm \hbar\omega_{\mathbf{q}})~. \label{eq:theo_scattering_transition_ep_nd}
\end{align}
For both electron-electron and electron-phonon interaction, the diagonal $(d)$ and non-diagonal $(nd)$ scattering rates can be interpreted as the mean value of in- and out- scattering rates of the electron occupations, Eq.~(\ref{eq:theo_scattering_occupation_ee_out}-\ref{eq:theo_scattering_occupation_ep_in}), summed over both bands. Analogously to the in- and out-scattering terms for the electron occupations, the diagonal and non-diagonal rates, respectively, account for decay and build-up of the coherent interband transition $p_\mathbf{k}^{vc\sigma}$. 
In addition, scattering completely non-linear in the optical excitation occurs included in $\Gamma_{\mathbf{k}}^{\lambda\sigma,nl}$ and $\Gamma_{\mathbf{k}}^{p\sigma,nl}$, as well as processes relying exclusively on exchange contributions in $T_{\mathbf{k}}^{p\sigma,ee}$. Since we will focus in this paper on the linear optical response, they are neglected in Sec.~\ref{sec:results}, but shown for completeness to App.~\ref{app:scattering}.
\\

The two-band  MBBE in Eqs.~(\ref{eq:theo_background:boltzmann}, \ref{eq:theo_background:bloch}, \ref{eq:theo_scattering_occupation_scattering}, \ref{eq:theo_scattering_transition_scattering}) resemble the SBE for a fully occupied valence band and a partially occupied conduction band. They represent a set of coupled many-body equations, providing the footing for a microscopic description of non-equilibrium electron dynamics for optically and transport induced intra- and interband processes in noble metal systems. For example, they extend single band plasmon decay into hot electrons by including interband transitions as additional decay channel \cite{bernardi_theory_2015}. In this way, the MBBE describe temporally and spectrally resolved linear and non-linear optical processes.\\
As a first example for their application, we calculate linear optical spectra for gold bulk material and compare with experimental data. To derive explicit band structure parameters for gold, we apply, in the following, an anisotropic band structure model \cite{rosei_d_1973}.

\section{\label{sec:dispersion_relation} Anisotropic Band Structure Model}
\begin{figure*}
    \centering
    \includegraphics[width=0.99\linewidth]{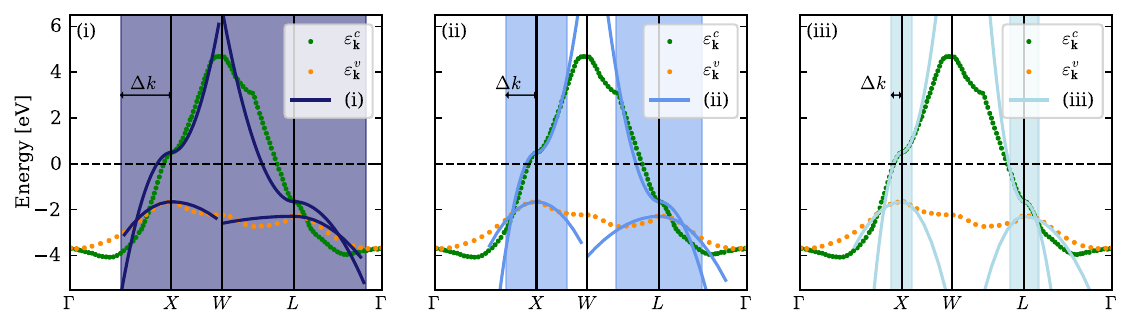}
    \caption{\justifying The electronic band structure of the $5d$ valence- and $6sp$ conduction band in gold, calculated  \textit{ab initio} by Rangel et al. \cite{rangel_band_2012}  (dotted lines in (i-iii)), is quadratically expanded in the vicinities of the $X$ and $L$ high symmetry points in both directions towards the $\Gamma$ ($k_{\parallel}$) and the $W$ point ($k_{\perp}$). Three different wave number ranges of optimal fit $\Delta k$, indicated by the shaded areas, are compared. In (i) we consider the entirety of the dispersion along the $PW$ paths ($P \in \{X,L\}$) such that it holds $\Delta k = \overline{PW}$. This allows for the fit to be in good global agreement with the dispersion. In (ii) we choose $\Delta k = 0.6\overline{PW}$ and in (iii) $\Delta k = 0.2\overline{PW}$, resulting in a more accurate fit in the vicinities of the $X$ and $L$ points at the cost of a worse overall agreement with the band structure.} 
    \label{fig_dispersion}
\end{figure*}

To model the highest valence and lowest conduction band in noble metals, we use the anisotropic band structure model introduced by Ref.~\cite{rosei_d_1973, guerrisi_splitting_1975}.
This approach expands the valence- and conduction band in noble metals quadratically in the vicinities of the six $X-$ and eight $L-$high symmetry points of the fcc Brillouin zone in radial direction towards the $\Gamma$ point and tangentially towards the $W$ point, cf.~Fig.~\ref{fig:theo:dispersion_skizze}. In Fig.~\ref{fig_dispersion}, this expansion is done for three ranges of optimal fit of \textit{ab initio} data provided in Fig.~8 of Ref.~\cite{rangel_band_2012}. These ranges are indicated by the shaded areas in Fig.~\ref{fig_dispersion}(i-iii), taking into account either the global band curvature, Fig.~\ref{fig_dispersion}(i), or the local band curvature in the immediate vicinity of high symmetry points, Fig.~\ref{fig_dispersion}(iii). 
The impact of the band structure fit parameters on the calculated optical spectra is discussed in Sec.~\ref{sec:results}.
\begin{table}[h]
    \caption{\label{tab:anisotropic_fits_i}%
    Fit parameters of the anisotropic dispersion model in Fig.~\ref{fig_dispersion}(i), $\Delta k = \overline{PW}$}\begin{ruledtabular}
        \begin{tabular}{ccccc}
            Band $\lambda$ & Point $P$ & $m^{\lambda P}_{\parallel}$  & $m^{\lambda P}_{\perp}$ & $\epsilon^{\lambda P}$ \\
            \hline
            $c$ & $X$ & -0.36 $m_e$ & 0.35 $m_e$  & 0.50 eV\\
            $c$ & $L$ & -1.21 $m_e$ & 0.54 $m_e$  & -1.65 eV\\
            $v$ & $X$ & -1.34 $m_e$ & -2.45 $m_e$  & -1.65 eV\\
            $v$ & $L$ & -2.18 $m_e$ & -14.6 $m_e$  & -2.27 eV\\
        \end{tabular}
    \end{ruledtabular}
\vspace{0.5cm}
    \caption{\label{tab:anisotropic_fits_ii}%
    Fit parameters of the anisotropic dispersion model in Fig.~\ref{fig_dispersion}(ii), $\Delta k = 0.6\overline{PW}$ }\begin{ruledtabular}
        \begin{tabular}{ccccc}
            Band $\lambda$ & Point $P$ & $m^{\lambda P}_{\parallel}$  & $m^{\lambda P}_{\perp}$ & $\epsilon^{\lambda P}$ \\
            \hline
            $c$ & $X$ & -0.19 $m_e$ & 0.25 $m_e$  & 0.50 eV\\
            $c$ & $L$ & -0.50 $m_e$ & 0.28 $m_e$  & -1.65 eV\\
            $v$ & $X$ & -1.01 $m_e$ & -0.98 $m_e$  & -1.65 eV\\
            $v$ & $L$ & -1.39 $m_e$ & -2.56 $m_e$  & -2.27 eV\\
        \end{tabular}
    \end{ruledtabular}
\vspace{0.5cm}
    \caption{\label{tab:anisotropic_fits_iii}%
    Fit parameters of the anisotropic dispersion model in Fig.~\ref{fig_dispersion}(iii), $\Delta k = 0.2\overline{PW}$ .}\begin{ruledtabular}
        \begin{tabular}{ccccc}
            Band $\lambda$ & Point $P$ & $m^{\lambda P}_{\parallel}$  & $m^{\lambda P}_{\perp}$ & $\epsilon^{\lambda P}$ \\
            \hline
            $c$ & $X$ & -0.11 $m_e$ & 0.17 $m_e$  & 0.50 eV\\
            $c$ & $L$ & -0.16 $m_e$ & 0.12 $m_e$  & -1.65 eV\\
            $v$ & $X$ & -0.99 $m_e$ & -0.44 $m_e$  & -1.65 eV\\
            $v$ & $L$ & -1.17 $m_e$ & -0.42 $m_e$  & -2.27 eV\\
        \end{tabular}
    \end{ruledtabular}
\end{table}

\begin{figure}
    \begin{subfigure}[t]{0.7\linewidth}
        \includegraphics[width=\linewidth]{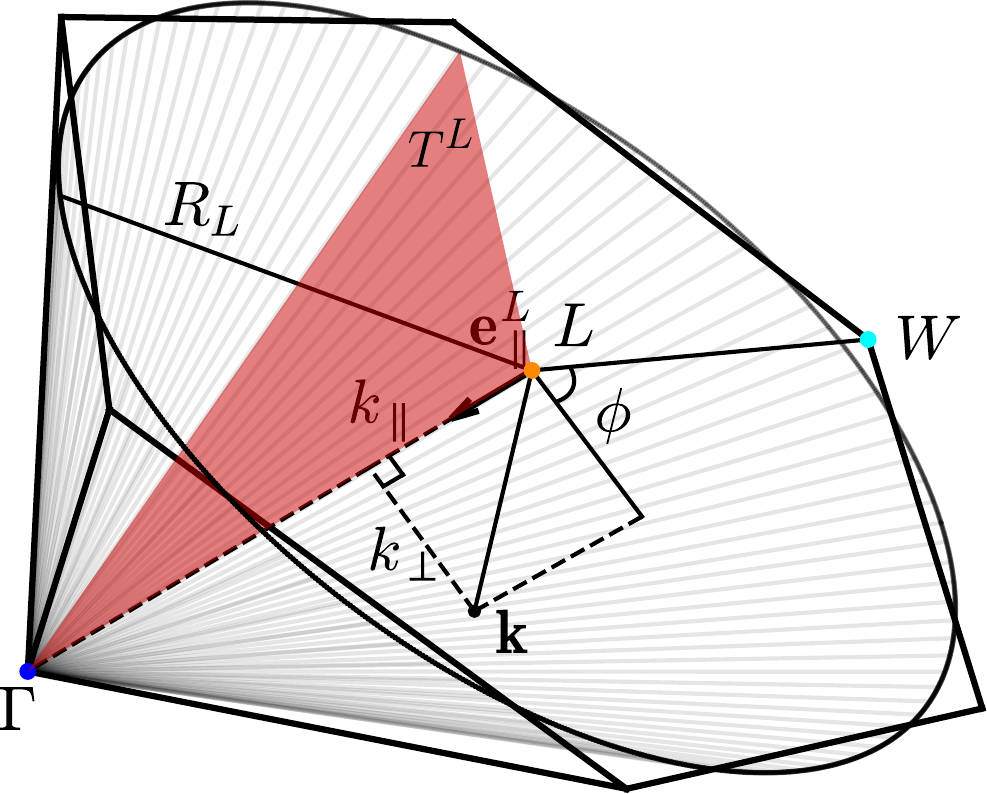}
        \vspace{6pt}
    \end{subfigure}
\caption{\justifying The first Brillouin zone is approximated by fourteen high symmetry cones $C^P$ with circular base of radius $R_P$ (shown here for one hexagonal $L$-face of the Brillouin zone), such that the area of the circular base is equal to the area of each face. The coordinates of the momentum $\mathbf{k} = (k_{\parallel},k_{\perp},\phi)$ are defined in Eq.~(\ref{eq:dispersion_coordinates}), relative to the $X$ or $L$ high symmetry point. The rotation of the red shaded triangle $T^{L}$ around the $\Gamma L$ axis generates the cone $C^L$.}
\label{fig_coordinates}
\end{figure}

\begin{figure}
    \begin{subfigure}[t]{\linewidth}
        \includegraphics[width=\linewidth]{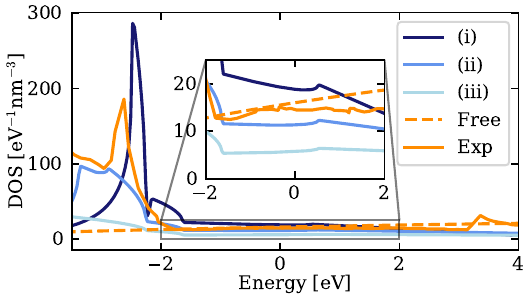}
    \end{subfigure}
\caption{\justifying The electronic DOS of the anisotropic two-band model for the wave number ranges of optimal fit in Fig.~\ref{fig_dispersion}(i-iii) is compared to the free electron DOS as well as to experimental data from Smith et al. \cite{smith_photoemission_1974}. The inset shows a zoom-in for energies relevant for this paper.}
\label{fig_dos}
\end{figure}

By approximating the Brillouin zone with cones with circular bases which open to the $X$- and  $L$-high symmetry points \cite{ziman_ordinary_1961}, cp. Fig.~\ref{fig_coordinates}, the momentum vector $\mathbf{k}$ is expressed in cylindrical coordinates and decomposed in a radial ($k_\parallel$) and two tangential ($k_\perp,\phi$) components originated at the corresponding high symmetry point $P \in \{X,L\}$ \cite{rosei_d_1973,guerrisi_splitting_1975}. These coordinates are defined as
\begin{align}
    k_{\parallel} = \mathbf{k}\cdot \mathbf{e}_{\parallel}^{P},\qquad k_{\perp} = \norm{\mathbf{k} - k_{\parallel}\mathbf{e}_{\parallel}^{P}},\label{eq:dispersion_coordinates}
\end{align}
with the radial unit vector $\mathbf{e}_{\parallel}^{P} = {\overline{P\Gamma }}/\norm{\overline{P\Gamma }}$ pointing towards the $\Gamma$-point.
 The angle $\phi$ can be arbitrarily chosen. Hereby, the condition $\frac{k_{\perp}}{R^P} \geq \frac{k_{\parallel}}{\overline{\Gamma P}}$ must hold for the coordinates to lie within a given cone, with $R^P$ the radius of a cones base.
With this we apply for both bands a quadratic dispersion in radial and tangential direction \cite{rosei_d_1973}
\begin{align}
    \varepsilon_{k_{\perp}k_{\parallel}}^{\lambda P} &=  \varepsilon^{\lambda P} +\frac{\hbar^2}{2m_{\perp}^{\lambda P}}k_{\perp}^{2} + \frac{\hbar^2}{2m_{\parallel}^{\lambda P}}k_{\parallel}^{2}~. \label{eq:dispersion:aniso_dispersion}
\end{align}
The parameters occurring in Eqs.~(\ref{eq:dispersion:aniso_dispersion}) are the effective masses $m_{\perp}^{\lambda P}$ and $m_{\parallel}^{\lambda P}$ in radial and tangential directions, respectively, and the energy offset $\varepsilon^{\lambda P}$ at the high symmetry points against the Fermi energy $E_F=0$.
\\
All twelve parameters $\{m_{\perp}^{\lambda P}, m_{\parallel}^{\lambda P}, \varepsilon^{\lambda P}\}$ at the symmetry points $P \in {X,L}$ are provided in Tab.~\ref{tab:anisotropic_fits_i}-\ref{tab:anisotropic_fits_iii} for the three fit ranges (i-iii) displayed in Fig.~\ref{fig_dispersion}. 
The density of states (DOS) for all ranges (i-iii) along with experimental data \cite{smith_photoemission_1974} is shown in  Fig.~\ref{fig_dos}. The van Hove singularities of the valence band are qualitatively reproduced by the fit range (i).  The Fermi surface yielded by the fit range in (i) is shown in Fig.~\ref{fig:theo:dispersion_skizze}, it features the typical necks at the $L$ points, as well as the domes at the $X$ points.\\
After the derivation of the susceptibility in Sec.~\ref{sec:linearization}, the effect of the anisotropic model for $\varepsilon_{k_{\perp}k_{\parallel}}^{\lambda P}$, Eq.~(\ref{eq:dispersion:aniso_dispersion}), on the optical response will be discussed in Sec.~\ref{sec:results}.

\section{\label{sec:linearization} Application of the MBBE to linear spectra}
In order to study linear spectra of noble metals, where the equilibrium electron occupations are only weakly perturbed \cite{grumm_theory_2025}, we linearize the MBBE in  Eq.~(\ref{eq:theo_background:boltzmann},\ref{eq:theo_scattering_occupation_scattering}) and Eq.~(\ref{eq:theo_background:bloch},\ref{eq:theo_scattering_transition_scattering}), respectively.
Both the electron occupations and interband transitions are expanded up to a first order perturbation in the optical field $\mathbf{E}(t)$, i.e., 
\begin{align}
    f_{\mathbf{k}}^{\lambda\sigma}(t) &= f_{\mathbf{k}}^{\lambda\sigma,eq} + f_{\mathbf{k}}^{\lambda\sigma,1}(t)~,\\
    p_{\mathbf{k}}^{vc\sigma}(t) &= p_{\mathbf{k}}^{vc\sigma,eq} +  p_{\mathbf{k}}^{vc\sigma,1}(t).
\end{align}
In equilibrium, at temperature $T$, the initial electron occupations $f_{\mathbf{k}}^{\lambda\sigma,eq}$ follow the Fermi-Dirac distribution
\begin{align}
    f_{\mathbf{k}}^{\lambda\sigma,eq} = \{\exp((\varepsilon_{\mathbf{k}}^{\lambda}-\mu)/k_BT) + 1\}^{-1}. \label{eq:linearization:fermi_distribution}
\end{align}
In the scope of this work, we approximate the chemical potential $\mu(T) \approx E_F$, which holds for all relevant equilibrium temperatures.\\
Since Eq.~(\ref{eq:linearization:fermi_distribution}) does not depend on spin, we omit the spin index $\sigma$ of the equilibrium electron occupations, $f_{\mathbf{k}}^{\lambda\sigma,eq} \rightarrow f_{\mathbf{k}}^{\lambda,eq}$. At this stage, no interband transitions exist without optical excitation, i.e $p_{\mathbf{k}}^{vc\sigma,eq} = 0 $. The dynamical first order perturbations $f_{\mathbf{k}}^{\lambda\sigma,1}(t),~ p_{\mathbf{k}}^{vc\sigma,1}(t)$ are induced by the optical excitation $\vb{E}(t)$.\\
Thus, the light-matter contributions from Eq.~(\ref{eq:theo_background:boltzmann},\ref{eq:theo_background:bloch}) take the form
\begin{align}
    \dot{f}_{\mathbf{k}}^{\lambda\sigma,1}\big\vert_{lm} &= \frac{e}{\hbar}\mathbf{E}(t)\cdot \nabla_{\mathbf{k}}f_{\mathbf{k}}^{\lambda,eq}~,\label{eq:lin:mbe_occupation}\\
    \dot{p}_{\mathbf{k}}^{vc\sigma,1}\big\vert_{lm} &= \frac{i}{\hbar}\{(\varepsilon_{\mathbf{k}}^v  -\varepsilon_{\mathbf{k}}^{c})p_{\mathbf{k}}^{vc\sigma,1}+ \mathbf{E}(t)\cdot\mathbf{d}_{\mathbf{k}}^{vc}(f_{\mathbf{k}}^{v,eq} - f_{\mathbf{k}}^{c,eq}) \}~.\label{eq:lin:mbe_transition}
\end{align}
The linearization of the mean field contributions from Eq.~(\ref{eq:theo:mean_field_occupations},\ref{eq:theo:mean_field_polarizations}) yield
\begin{align}
    \dot{f}_{\mathbf{k}}^{\lambda\sigma,1}\big\vert_{HF} &= 0~,\label{eq:lin:mean_field_occupations}\\
    \dot{p}_{\mathbf{k}}^{vc\sigma,1}\big\vert_{HF} &= \frac{i}{\hbar} \sum_{\vb{qG}}(V_{\vb{q}}^{vc}f_{\mathbf{k}}^{v\sigma,eq} - V_{\vb{q}}^{cv}f_{\vb{k}}^{c\sigma,eq})p_{\vb{k+q+G}}^{vc\sigma,1}(t)~.\label{eq:lin:mean_field_polarizations}
\end{align} 
Eq.~(\ref{eq:lin:mean_field_polarizations}) leads to a renormalization of the light-matter coupling $\mathbf{E}(t)\cdot \mathbf{d}_{\mathbf{k}}^{vc}$ and constitutes a self consistency problem in the first order of the interband transitions $p_{\mathbf{k+q}}^{vc\sigma,1}$. 
In semiconductors, this leads to the formation of excitons, but is of transient character \cite{schone_transient_2002} and thus negligible in metals due to the strong screening and lack of a band gap.

\subsection{Linearized Scattering Rates}

The linearization of the scattering  terms in the electron occupation dynamics in Eq.~(\ref{eq:theo_scattering_occupation_scattering}) results in the expression
\begin{align}
    \dot{f}_{\mathbf{k}}^{\lambda\sigma,1}\big\vert_{sc} &= -(\Gamma_{\mathbf{k}}^{\lambda\sigma,out,eq} + \Gamma_{\mathbf{k}}^{\lambda\sigma,in,eq})f_{\mathbf{k}}^{\lambda\sigma,1} \label{eq:lin:occupation_rta} \\
    &\qquad \qquad - \Gamma_{\mathbf{k}}^{\lambda\sigma,out,1}f_{\mathbf{k}}^{\lambda\sigma,eq} + \Gamma_{\mathbf{k}}^{\lambda\sigma,in,1}(1-f_{\mathbf{k}}^{\lambda\sigma,eq})~.\nonumber
\end{align}
This is summarized into an exponential decay rate $\gamma_\mathbf{k}^\lambda$ of the optically induced perturbations $f_{\mathbf{k}}^{\lambda\sigma,1}$,
\begin{align}
    \dot{f}_{\mathbf{k}}^{\lambda\sigma,1}\big\vert_{sc} = -(\gamma_{\mathbf{k}}^{\lambda,ep} + \gamma_{\mathbf{k}}^{\lambda,ee})f_{\mathbf{k}}^{\lambda\sigma,1} = -\gamma_{\mathbf{k}}^{\lambda}f_{\mathbf{k}}^{\lambda\sigma,1}~. \label{eq:lin:occupation_gamma}
\end{align}

\begin{table}
    \caption{\label{tab:parameters}%
    Numerical material parameters for gold.}\begin{ruledtabular}
        \begin{tabular}{lcr}
            Parameter & Value & Reference \\
            \hline
            $n_0$ & $59.0~\frac{1}{\text{nm}^3} $ & \cite{ashcroft_solid_1976} \\
            $a_0$ & $0.408$~nm & \cite{ashcroft_solid_1976} \\
            $E_F$ & $5.53 $~eV & \cite{ashcroft_solid_1976} \\
            $k_F $ & $ 12.0~\frac{1}{\text{nm}}$ & from $E_F= \frac{\hbar^2 k_F^2}{2m}$ \\
            $q_{D}$ & $15.2~\frac{1}{\text{nm}} $ & from $q_{D} = (6\pi^2 n_0)^{\frac{1}{3}}$ \\
            $\kappa_{TF}$ & $16.9~\frac{1}{\text{nm}} $ & \cite{ashcroft_solid_1976}\\
            $c_{LA}$ & $3.24\cdot 10^ {-3}~\frac{\text{nm}}{\text{fs}} $ & \cite{haynes_crc_2011} \\
            $m^c$ & $0.99~m_e$ & \cite{ashcroft_solid_1976}\\
            $m^v$ & $3.37~m_e$ & \cite{brown_ab_2016}\\
            $\varepsilon^c$ & $-E_F$ & \cite{ashcroft_solid_1976}\\
            $\varepsilon^v$ & $-3.7$eV & \cite{rangel_band_2012}
        \end{tabular}
    \end{ruledtabular}
\end{table}

The derivation of both electron-phonon $\gamma_{\mathbf{k}}^{\lambda,ep}$ and electron-electron $\gamma_{\mathbf{k}}^{\lambda,ee} $ relaxation rates follow a path provided in Ref.~\cite{grumm_theory_2025} and is discussed in App.~\ref{app:rra_full}. 
The linearization of the interband transition scattering equations in Eq.~(\ref{eq:theo_scattering_transition_scattering}) leads to a dephasing rates, when neglecting the non-diagonal contributions $\Gamma^{p\sigma,nd}_{\mathbf{kq}}$ and  exchange contributions $T_{\mathbf{k}}^{p\sigma,ee}$ in Eq.~(\ref{eq:theo_scattering_transition_scattering}) in random phase approximation \cite{hess_maxwell-bloch_1996}
\begin{align}
    \dot{p}_{\mathbf{k}}^{vc\sigma,1}\big\vert_{sc} & = -(\gamma_{\mathbf{k}}^{p,ep} + \gamma_{\mathbf{k}}^{p,ee})p_{\mathbf{k}}^{vc\sigma,1} = -\gamma_{\mathbf{k}}^{p}p_{\mathbf{k}}^{vc\sigma,1}~. \label{eq:lin:transition_rta}
\end{align}
In this approximation, the dephasing rates $\gamma_{\mathbf{k}}^{p}$ correspond to the linearized diagonal scattering rates $\Gamma_{\mathbf{k}}^{p\sigma,d}$ in Eq.~(\ref{eq:theo_scattering_transition_ee_d},\ref{eq:theo_scattering_transition_ep_d}) with $f_{\mathbf{k}}^{\lambda\sigma}\rightarrow f_{\mathbf{k}}^{\lambda,eq}$.\\
For a feasible numerical evaluation of relaxation $\gamma_{\mathbf{k}}^{\lambda}$ and dephasing rates $\gamma_{\mathbf{k}}^{p}$ in Eqs.~(\ref{eq:lin:occupation_gamma},\ref{eq:lin:transition_rta}), we apply an isotropic free electron dispersion \cite{ashcroft_solid_1976}
\begin{align}
    \varepsilon^{\lambda}_{\mathbf{k}} = \varepsilon^{\lambda} + \frac{\hbar^2}{2m^{\lambda}}|\mathbf{k}|^2, \label{eq:dispersion:dispersion_iso}
\end{align}
with isotropic effective masses $m^\lambda$ and energy offsets $\varepsilon^\lambda$ relative to the Fermi energy. The parameters are provided in Tab.~\ref{tab:parameters}. For the estimation of relaxation and dephasing rates, this dispersion model is justified, as the rates predominantly describe a redistribution of large momentum across the Brillouin zone \cite{grumm_femtosecond_2025}, therefore details of energy surfaces are of lesser importance.
Thus, in anisotropic coordinates $(k_{\perp},k_{\parallel},P)$ the relaxation and dephasing rates are given by 
\begin{align}
    \gamma_{\vb{k}}^{\lambda/p} \rightarrow \gamma_{P-k_{\parallel}}^{\lambda/p}~,    \label{eq:lin:conversion}
\end{align}
where $P-k_{\parallel}$ designates the radial momentum component relative to the center of the Brillouin zone.
\\
The electron-phonon relaxation rates in Eqs.~(\ref{eq:lin:occupation_gamma},\ref{eq:lin:transition_rta}) are divided into normal- ($N$) and Umklapp ($U$) processes, i.e $\gamma_{k}^{\lambda/p,ep} = \gamma_{k}^{\lambda/p,ep,N} + \gamma_{k}^{\lambda/p,ep,U}$ with
\begin{widetext}
    \begin{align}
    \gamma_{k}^{\lambda,ep,N} &= -\frac{Vm^{\lambda}}{2\pi k_{B}T\hbar^3 k^3}\int_0^{q_D}dq q^3n(\hbar\omega_q)n(-\hbar \omega_q)|g_{q0}^\lambda|^2\hbar\omega_q , \label{eq:lin:occupation_normal_rate_ep}\\
    \gamma_{k}^{\lambda,ep,U} &= -\frac{Vm^{\lambda}}{2\pi k_B T  \hbar^3 k^3}\int_{2(q_D - k)}^{q_D}dq q n(\hbar\omega_q)n(-\hbar\omega_q)
    |g_{q\hat{G}}^{\lambda}|^2\hbar\omega_q (4q_D^2 - q^2), \numberthis \label{eq:lin:occupation_umklapp_rate_ep}
\end{align}
\begin{align}
    &\gamma_{k}^{p,ep,N} = \sum_{\lambda} \frac{V|m^{\lambda}|}{4\pi\hbar^3 k} \int_{0}^{q_D}dq q |g_{q0}^{\lambda}|^2\{2n(\hbar\omega_q) + 1 
   - \frac{2}{k_B T}f_k^{\lambda,eq}(1-f_k^{\lambda,eq})\hbar\omega_q\},\label{eq:lin:transition_normal_rate_ep}\\
    &\gamma_{k}^{p,ep,U} = \sum_{\lambda}\frac{V|m^{\lambda}|(2q_D - k)}{4\pi \hbar^3 k^2}\int_{2(q_D - k)}^{q_D}dq q|g_{q\hat{G}}^{\lambda}|^2 \{2n(\hbar\omega_q) + 1 - \frac{2}{k_B T}f_k^{\lambda,eq}(1-f_k^{\lambda,eq})\hbar\omega_q\}.\label{eq:lin:transition_umklapp_rate_ep}
\end{align}
Here $n(\hbar\omega_q) = \{\exp(\frac{\hbar\omega_q}{k_B T})-1\}^{-1}$ designates the Bose-Einstein distribution, $q_D = (6\pi n_i)^{\frac{1}{3}}$ the Debye wave number, and $n_i$ the ion density which coincides for noble metals with one conduction band electron per atom with the electron density $n_0$, cp. Tab.~\ref{tab:parameters}. Furthermore holds $\hat{G} = \{k\frac{4q_D^2 - q^2}{2q_D-k}\}^{1/2}-q$. We refer to App.~\ref{app:rta} for insight on the derivation of Eqs.~(\ref{eq:lin:occupation_normal_rate_ep}-\ref{eq:lin:transition_umklapp_rate_ep}).\\
The rates describing electron-phonon normal processes in Eq.~(\ref{eq:lin:occupation_normal_rate_ep},\ref{eq:lin:transition_normal_rate_ep}) account for all quasi-momentum conserving scattering processes as no reciprocal lattice vectors are involved. For the relaxation rate in Eq.~(\ref{eq:lin:occupation_normal_rate_ep}) the expression corresponds to the well known Bloch-Grüneisen formula \cite{czycholl_solid_2023,ashcroft_solid_1976}.
The rates involving umklapp processes in Eq.~(\ref{eq:lin:occupation_umklapp_rate_ep},\ref{eq:lin:transition_umklapp_rate_ep}) consider quasi-momentum non-conserving processes.\\
For the electron-electron scattering, the relaxation and dephasing rates are split into four parts, resulting from the cases of two electrons respectively either undergoing a normal or an Umklapp process, $\gamma_{k}^{\lambda/p,ee} = \gamma_{k}^{\lambda/p,ee,NN} + \gamma_{k}^{\lambda/p,ee,NU} +\gamma_{k}^{\lambda/p,ee,UN} +\gamma_{k}^{\lambda/p,ee,UU}$. For the intraband dynamics,  $\gamma_{k}^{c,ee,NN} $ vanishes, as the relaxation of the momentum-polarized electron system can proceed only via quasi-momentum non-conserving processes  \cite{lawrence_umklapp_1972}. The remaining electron-electron rates read

\begin{align}
   \gamma_{k}^{c,ee,NU} = \frac{V^2 m_c^2q_D}{4\pi^3\hbar^5 kk_Fk_B T} \int_{2(q_D-k_F)}^{q_D+k}\int_{-1}^{p_M} dq  dp q^2\Phi_{kqp}^{c,N}  p\{\frac{4q_D^2 - 4q_D k_F}{q^2} + 1\},\label{eq:lin:gamma_occupation_NU}
\end{align}

\begin{align}
    \gamma_{k}^{c,ee,UN}  = \frac{-V^2 m_c^2q_D}{2\pi^3\hbar^5 kk_B T}\int_{q_D - k}^{2k_F}\int_{p_m}^{1}dqdp q \Phi_{kqp}^{c,U} \frac{k + qp}{\sqrt{k^2 + q^2 + 2kqp}},\label{eq:lin:gamma_occupation_UN}
\end{align}

\begin{align}
   &\gamma_{k}^{c,ee,UU} = \frac{V^2 m_c^2q_D}{4\pi^3\hbar^5 kk_Fk_B T}\int_{2(q_D - k_F)}^{2q_D}\int_{p_m}^{1}dqdp q^2 \Phi_{kqp}^{c,U}
    \{p(\frac{4q_D^2 - 4q_D k_F}{q^2} + 1) -2\frac{k+qp}{\sqrt{k^2 + q^2 + 2kqp}}\frac{2q_D - k_F}{q}\} ,\label{eq:lin:gamma_occupation_UU}
\end{align}

\begin{align}
    \gamma_{k}^{p,ee,NN} + \gamma_{k}^{p,ee,NU} = &\frac{-V^2 m_c^2}{8\pi^3\hbar^5 k_Fk_B T}\sum_{\lambda}\Big\{k_F\int_{0}^{2k_F}\int_{-1}^{p_M}dq dp q \Phi_{kqp}^{\lambda,N}
     +(2q_D -k_F)\int_{2(q_D - k)}^{q_D + k}\int_{-1}^{p_M}dq dp q \Phi_{kqp}^{\lambda,N}\Big\}, \label{eq:lin:gamma_polarization_N}
\end{align}

\begin{align}
    \gamma_{k}^{p,ee,UN} + \gamma_{k}^{p,ee,UU} = &\frac{-V^2 m_c^2}{8\pi^3\hbar^5 k_Fk_B T}\sum_{\lambda}\Big\{k_F\int_{q_D - k}^{2k_F}\int_{p_m}^{1}dq dp q \Phi_{kqp}^{\lambda,U}
     +(2q_D -k_F)\int_{2(q_D - k_F)}^{2q_D }\int_{p_m}^{1}dq dp q \Phi_{kqp}^{\lambda,U}\Big\}, \label{eq:lin:gamma_polarization_U}
\end{align}
with 
\begin{align}
   \Phi_{kqp}^{\lambda,N/U} &= |V_{q}^{c\lambda}|^2n(\Delta E_{kqp}^{\lambda,N/U})n(-\Delta E_{kqp}^{\lambda,N/U})(\Delta E_{kqp}^{\lambda,N/U})^2, \\
   \Delta E_{kqp}^{c,N} &= \frac{\hbar^2}{2m_c}(2kqp + q^2), \qquad \Delta E_{kqp}^{c,U} = \frac{\hbar^2}{2m_c}\{2kqp  + q^2 -  4q_D\sqrt{k^2 + q^2 + 2kqp} +  4q_D^2\}.
\end{align}  

\end{widetext}

The limits $p_m := \max(\frac{q_D^2 - k^2 - q^2}{2kq},-1)$ and $p_M :=  \min(\frac{q_D^2 - k^2 - q^2}{2kq},1)$ ensure the distinction between normal and Umklapp processes.\\

To keep the evaluation of the rates in Eq.~(\ref{eq:lin:gamma_occupation_NU}-\ref{eq:lin:gamma_polarization_U}) clear, we assume that the direct and exchange in electron-electron scattering in Eqs.~(\ref{eq:theo_scattering_occupation_ee_out},\ref{eq:theo_scattering_occupation_ee_in},\ref{eq:theo_scattering_transition_ee_d},\ref{eq:theo_scattering_transition_ee_nd}) for same band and spin cancel out ($V_{\mathbf{q}}^{\lambda\lambda}-V_{\mathbf{k'-q-k}}^{\lambda\lambda} \approx 0 $), which is an appropriate approximation for high carrier, spin-randomized, densities \cite{moskova_exchange_1994}. This reproduces the standard form of Boltzmann collision approaches \cite{del_fatti_nonequilibrium_2000,rethfeld_ultrafast_2002,mueller_relaxation_2013}. \\
We refer again to App.~\ref{app:rta} for insight on the derivation of the electron-electron rates in Eq.~(\ref{eq:lin:gamma_occupation_NU}-\ref{eq:lin:gamma_polarization_U}).\\

\subsection{Linear Optical Response}
The linearized equations in Eq.~(\ref{eq:lin:mbe_occupation},\ref{eq:lin:occupation_rta}) for the electron occupations can be solved explicitly in frequency space in the coordinates of the anisotropic band structure model, in Sec.~\ref{sec:dispersion_relation},
\begin{align}
    \hat{f}_{k_{\perp}k_{\parallel}\phi}^{\lambda \sigma P,1}(\omega) = \frac{-e\hat{\mathbf{E}}(\omega)\cdot\nabla_{k_{\perp}k_{\parallel}\phi}f_{k_{\perp}k_{\parallel}}^{\lambda P,eq}}{i\hbar\omega -  \hbar\gamma_{P - k_{\parallel}}^{\lambda}}~.\label{eq:lin:occupation_explicit}
\end{align}
Here, the equilibrium occupations are independent of $\phi$. 
Since in the fully occupied valence band holds $f_{k_{\perp}k_{\parallel}}^{v P ,eq} = 1$, the momentum gradient vanishes, $\nabla_{k_{\perp}k_{\parallel}\phi}f_{k_{\perp}k_{\parallel}}^{vP,eq}=0$, and the discussion of intraband dynamics remains only  on the partially filled conduction band.
With the macroscopic current density in Eq.~(\ref{eq:theo:p_intra}) and the linearized electron occupations in Eq.~(\ref{eq:lin:occupation_explicit}), we retrieve the current density in frequency space
\begin{align}
    &\mathbf{j}(\omega) =  -i\omega\varepsilon_0 \overline{\overline{\chi}}_{\text{intra}}(\omega)\hat{\mathbf{E}}(\omega)\nonumber \\
    =&\frac{2e^2}{V k_B T }\sum_{P,\mathbf{k}\in C^P}\frac{\mathbf{v}_{k_{\perp}k_{\parallel}\phi}^{cP}\otimes\mathbf{v}_{k_{\perp}k_{\parallel}\phi}^{cP} f_{k_{\perp}k_{\parallel}}^{cP,eq}(1-f_{k_{\perp}k_{\parallel}}^{cP,eq})}{i\omega - \gamma_{P - k_{\parallel}}^{c}}\hat{\vb{E}}(\omega). \label{eq:lin:j_explicit}
\end{align}
Here, we approximate the summation over the first Brillouin zone $\sum_{\mathbf{k}}$ by the summation over the union of all high symmetry point cones $C^P$, cp. Fig.~\ref{fig_coordinates} and use, that the zeroth order contribution of the electron occupations vanishes as a direct consequence of Kramer's theorem \cite{czycholl_basics_2023}. The prefactor of 2 stems from the summation over the spin degree of freedom. 
Even in the anisotropic model the tensor form of Eq.~(\ref{eq:lin:j_explicit}) can be reduced to a scalar expression, as the contributions of cones of one kind (either $X-$ or $L-$ cones)  turn out in a way such that in total the current density is parallel to the incident electrical field, cf. App.~\ref{app:dyadic}.
The resulting intraband susceptibility ultimately reads
\begin{align}
    \chi_{\text{intra}}(\omega) =\frac{2e^2}{3V\varepsilon_0 k_B T}\sum_{P,\mathbf{k}\in C^{P}}\frac{|\mathbf{v}_{k_{\perp}k_{\parallel}\phi}^{cP}|^2f_{k_{\perp}k_{\parallel}}^{cP,eq}(1-f_{k_{\perp}k_{\parallel}}^{cP,eq})}{\omega^2 + i\omega\gamma_{P - k_{\parallel}}^{c}}~. \label{eq:lin:chi_intra}
\end{align}
\\

\begin{figure*}[ht!]
    \centering
    \includegraphics[width=\linewidth]{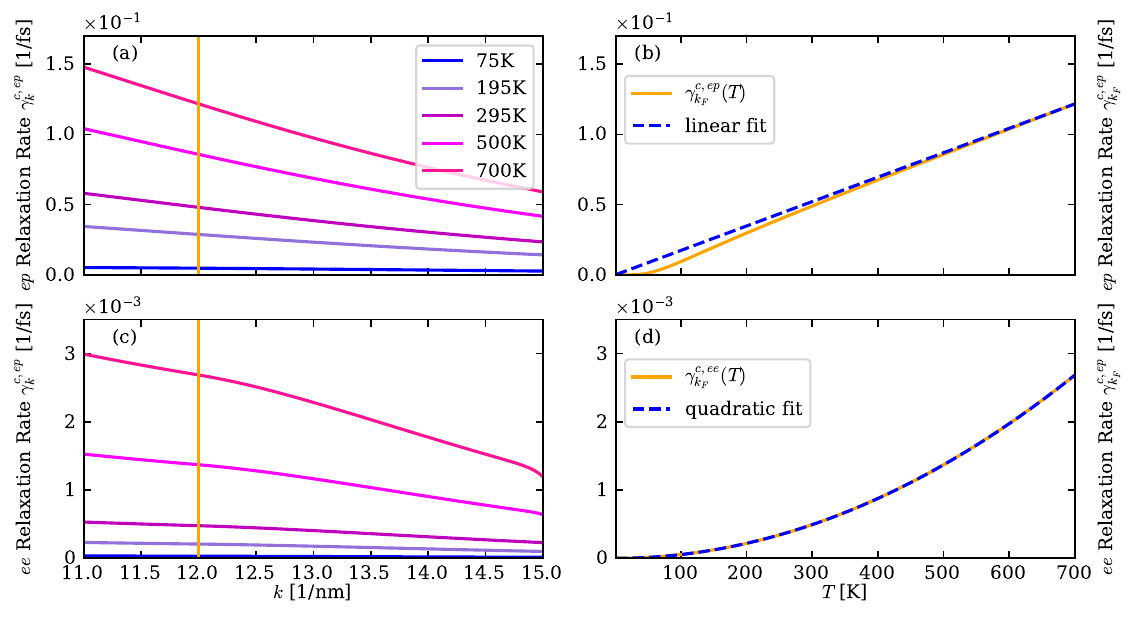}
    \caption{\justifying Temperature dependence of the (a,b) electron-phonon and (c,d) electron-electron  relaxation rates. In (a) and (c) the rates are shown from 75~K to 700~K as function of the wave number $k$. In (b) and (d) the rates are evaluated at the Fermi wavenumber $k_F=12.0~\text{nm}^{-1}$ and shown as a function of temperature (orange lines). The electron-phonon and electron-electron rates are respectively linearly and quadratically fitted (blue dashed lines).}
    \label{fig:relaxation_rates}
\end{figure*}

\begin{figure*}[ht!]
    \centering
    \includegraphics[width=\linewidth]{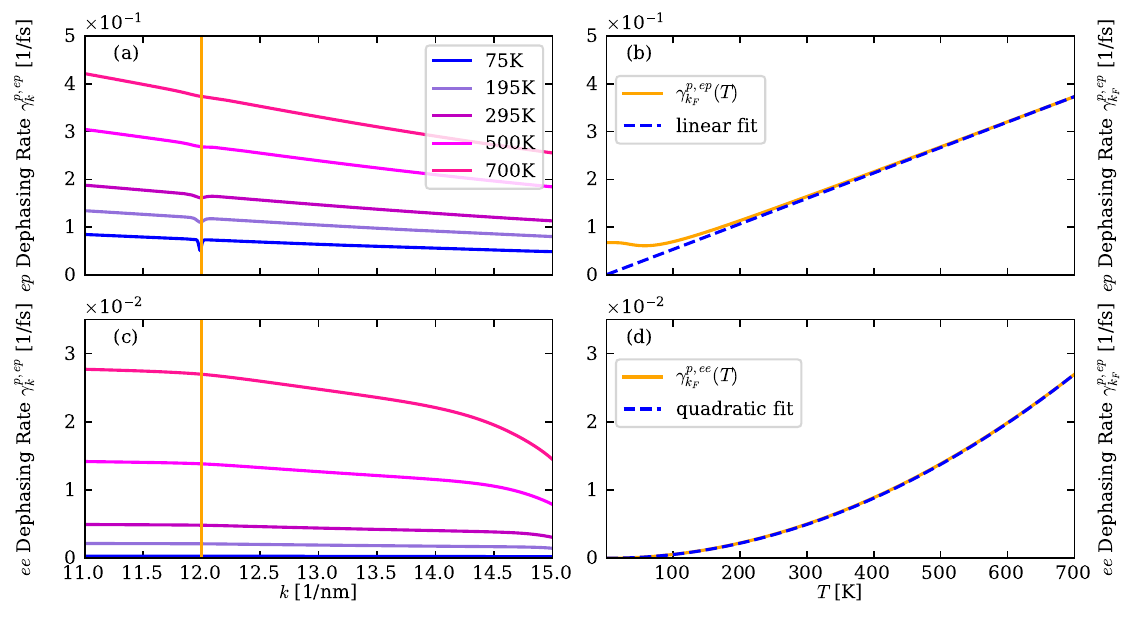}
    \caption{\justifying Temperature dependence of the (a,b) electron-phonon and (c,d) electron-electron dephasing rates. In (a) and (c) the rates are shown from 77~K to 900~K as function of $k$. In (b) and (d) the rates are evaluated at the Fermi wavenumber $k_F=12.0~\text{nm}^{-1}$ and shown as a function of temperature (orange line). The electron-phonon and electron-electron rates are respectively linearly and quadratically fitted (blue dashed lines).}
    \label{fig:dephasing_rates}
\end{figure*}

Similarly, with equations Eq.~(\ref{eq:lin:mbe_transition},\ref{eq:lin:transition_rta}) the interband transitions are solved in frequency domain
\begin{align}
    \hat{p}_{k_{\perp}k_{\parallel}\phi}^{vc\sigma P,1}(\omega) = \frac{-\hat{\mathbf{E}}(\omega)\cdot \mathbf{d}_{k_{\perp}k_{\parallel}\phi}^{vcP}(f_{k_{\perp}k_{\parallel}}^{vP,eq} - f_{k_{\perp}k_{\parallel}}^{cP,eq})  }{\hbar\omega + \varepsilon_{k_{\perp}k_{\parallel}}^{vP} -\varepsilon_{k_{\perp}k_{\parallel}}^{cP} + i\hbar\gamma_{P-k_{\parallel}}^{p}}. \label{eq:lin:transition_explicit}
\end{align} 
It should be noted that in the dephasing rate approximation, $p_{k_{\perp}k_{\parallel}\phi}^{vc\sigma} \neq p_{k_{\perp}k_{\parallel}\phi}^{cv\sigma *}$ are no longer hermitian. 
For simplicity, we consider momentum independent dipole matrix elements $\mathbf{d}_{k_{\perp}k_{\parallel}\phi}^{vcP} \rightarrow \mathbf{d}^P$, with $\mathbf{d}^{P} \parallel \vb{e}_{\parallel}^P$. 
With the macroscopic polarization in Eq.~(\ref{eq:theo:p_inter}) and Eq.~(\ref{eq:lin:transition_explicit}), we find
\begin{align}
    &\vb{P}_{\text{inter}}(\omega) = \varepsilon_0\chi_{\text{inter}}(\omega)\hat{\mathbf{E}}(\omega)~,\\
     &\mathbf{\chi}_{\text{inter}}(\omega) =  \label{eq:interband_susceptibility}
    \frac{2}{3V\varepsilon_0}\sum_{P,\mathbf{k}\in C^P}|\mathbf{d}^{P}|^2(f_{k_{\perp}k_{\parallel}}^{cP,eq} - f_{k_{\perp}k_{\parallel}}^{vP,eq})\\
    &\quad \frac{2(\varepsilon_{k_{\perp}k_{\parallel}}^{cP} - \varepsilon_{k_{\perp}k_{\parallel}}^{vP})}{(\hbar\omega)^2 - (\varepsilon_{k_{\perp}k_{\parallel}}^{cP} - \varepsilon_{k_{\perp}k_{\parallel}}^{vP})^2 + 2i\hbar\omega\hbar\gamma_{P-k_{\parallel}}^{p}  - (\hbar\gamma_{P-k_{\parallel}}^{p} )^2}~.\nonumber 
\end{align}
The prefactor of $\frac{2}{3}$ arises from the summation over the spin degree of freedom as well as from the angle averaging in the dyadic product, cp. App.~\ref{app:dyadic}.\\
The total permittivity is given by 
\begin{align}
    \epsilon(\omega) = \epsilon_{\infty} + \chi_{\text{intra}}(\omega) + \chi_{\text{inter}}(\omega)~, \label{eq:chi_tot}
\end{align}
with $\epsilon_{\infty}$ modeling a phenomenological dielectric background generated by the core electrons \cite{vial_improved_2005}.

\subsection{\label{sec:results:drude_lorentz}Comparison to Drude-Lorentz models}
 In plasmonics, a frequently used approach to the description of linear spectra of noble metals is the Drude-Lorentz model \cite{etchegoin_analytic_2006,pfeifer_time-domain_2024,pirzadeh_plasmoninterband_2014}. There, for modeling the interband susceptibility, a discrete set of parametrized effective transitions, indexed by $i$,  with resonances $\hbar\omega_i$, line widths $\Gamma_i$ and oscillator strengths $A_i$ is assumed,
 \begin{align}
     \chi_{\text{inter}}(\omega) = \sum_{i}\frac{A_i}{\omega^2 - \omega_{i}^2 + i\omega\Gamma_i}~.
 \end{align}
By identifying $i \rightarrow \{P,\mathbf{k}\in C^P\}$, with $\mathbf{k}$ the continuous momentum, it follows that
 \begin{align*}
     A_i &\equiv \frac{2}{3\hbar^2V\varepsilon_0}|\mathbf{d}^P|^2(f_{k_{\perp}k_{\parallel}}^{cP,eq}- f_{k_{\perp}k_{\parallel}}^{vP,eq})(\varepsilon_{k_{\perp}k_{\parallel}}^{cP} - \varepsilon_{k_{\perp}k_{\parallel}}^{vP} ),\\
     \omega_i^2 &\equiv (\varepsilon_{k_{\perp}k_{\parallel}}^{cP} - \varepsilon_{k_{\perp}k_{\parallel}}^{vP})^2/\hbar^2  + (\gamma_{P-k_{\parallel}}^p)^2 ,\\
     \Gamma_i &\equiv 2\gamma_{P-k_{\parallel}}^{p}.\numberthis\label{eq:results:micro_to_drude_lorentz}
 \end{align*}

 The most notable result of this comparison is the expression of phenomenological oscillator strength $A_i$ in terms of the microscopic occupations, dipole matrix elements and the momentum resolved band gap. \\
In the following section, Eqs.~(\ref{eq:lin:chi_intra},\ref{eq:interband_susceptibility}) are analyzed numerically.

\section{\label{sec:results} Results and Discussion} 
In this section we discuss the linear optical response of  bulk gold with material parameters listed in Tab.~\ref{tab:parameters}. We start with the numerical evaluations of electron-phonon and electron-electron relaxation and dephasing rates in Eqs.~(\ref{eq:lin:occupation_normal_rate_ep},\ref{eq:lin:transition_umklapp_rate_ep}) and Eqs.~(\ref{eq:lin:gamma_occupation_NU},\ref{eq:lin:gamma_polarization_U}), respectively, in Sec.~\ref{sec:results:rates}. With those rates we will then, in Sec.~\ref{sec:results:ranges_spectra}, discuss the influence of the fit parameters used to determine the anisotropic band model, cp. Fig.~\ref{fig_dispersion}, on the linear spectrum, Eqs.~(\ref{eq:lin:chi_intra},\ref{eq:interband_susceptibility}). The most promising of the three fit ranges will then be used in Sec.~\ref{sec:results:temperature_spectra} to investigate the temperature dependence of the linear spectra and the origin of the spectrally extended interband transition edge. These calculated spectra are compared with experimental data obtained in our own temperature-dependent ellipsometry experiments, cf.~App.~\ref{app:exp}.

\subsection{\label{sec:results:rates}Temperature dependent Relaxation- and Dephasing Rates}

The temperature dependence of the electron-phonon and electron-electron relaxation rates are shown in Fig.~\ref{fig:relaxation_rates}. From the quantitative comparison of Fig.~\ref{fig:relaxation_rates}(a) and \ref{fig:relaxation_rates}(b), it becomes clear that the electron-phonon rates dominate the intraband relaxation. Hereby, electron-phonon Umklapp processes, Eq.~(\ref{eq:lin:occupation_umklapp_rate_ep}), contribute the most, as the quasi momentum transfer for these processes is greater than for normal processes \cite{lawrence_umklapp_1972, grumm_theory_2025}.
In the high temperature regime, the electron-phonon relaxation rate, orange line in  Fig.~\ref{fig:relaxation_rates}(b), follows a linear temperature dependence (blue dashed line) while for low temperatures a $T^5$-dependence is obtained. This reflects the Bloch-Grüneisen relation for the electric resistance \cite{czycholl_solid_2023, bloch_zum_1930}. The electron-electron relaxation rate follow a $T^2$-law as in Fermi liquid theory \cite{kaveh_electron-electron_1984}.\\
For the dephasing of interband transitions, shown in Fig.~\ref{fig:dephasing_rates}, the electron-phonon rate in Fig.~\ref{fig:dephasing_rates}(a) provides the major contribution. However, here the normal and Umklapp contributions are similar and both exhibit a dip in the vicinity of the Fermi edge, particularly noticeable for lower temperatures. This is caused by the finite Pauli blocking terms $f_k^{c,eq}(1-f_k^{c,eq})$ in Eqs.~(\ref{eq:lin:transition_normal_rate_ep},\ref{eq:lin:transition_umklapp_rate_ep}) for conduction band electrons. This effectively leads to an extended lifetime of the interband transitions for momenta $k$ close to $k_F$. However, the underlying approximations causing this dip collapse for temperatures far below the Debye temperature. 
Nevertheless, the finite value of the electron-phonon dephasing rate in the low temperature limit, caused by phonon emission, is also observed in 2D semiconductors \cite{selig_excitonic_2016}.
\\
Similarly to intraband relaxation, the electron-phonon dephasing rate follows a linear temperature dependence for temperatures above the Debye temperature, as shown in Fig.~\ref{fig:dephasing_rates}(b). The electron-electron dephasing rate in Fig.~\ref{fig:dephasing_rates}(c, d) also follows a $T^2$ law.\\
It should be noted that the decline of both the electron-electron relaxation- and dephasing rates near the edge of the radius of the Brillouin zone $q_D$ is likely due to limits of the isotropic dispersion model.

\subsection{\label{sec:results:ranges_spectra} Linear Spectra as Function of the Dispersion Fit Range}

Based on the previously calculated relaxation and dephasing rates, we first discuss the linear spectra resulting from  Eqs.~(\ref{eq:lin:chi_intra},\ref{eq:interband_susceptibility}) at room temperature, $T = 295$~K, for the three fit ranges introduced in Fig.~\ref{fig_dispersion}. 
As only the occupations of states in the thermally broadened region close to the Fermi edge substantially deviate from the equilibrium distribution, the relaxation rate $ \gamma_{k_{\perp}k_{\parallel}}^{cP} \approx \gamma_{k_F}^{c}$ is approximated by a constant value at the Fermi edge $|\mathbf{k}| = k_F$ \cite{grumm_theory_2025}. The relaxation rate reflects the time scale of momentum-orientational relaxation in a linear regime, as discussed in \cite{grumm_femtosecond_2025,grumm_theory_2025,binder_greens_1997}, and represents the microscopic counterpart of the phenomenological macroscopic Drude collision rate \cite{drude_zur_1900}.
At room temperature, the inverse relaxation rate of $(\gamma_{k_F}^{c}({295\text{K}}))^{-1} = 20.8$ fs is in agreement with experimental \cite{olmon_optical_2012} and \textit{ab initio} studies \cite{brown_nonradiative_2016,mustafa_ab_2016}. Thus, the intraband susceptibility in Eq.~(\ref{eq:lin:chi_intra}) results in a Drude expression 
\begin{align}
    \chi_{\text{intra}}(\omega) = \frac{-\omega_{pl}^2}{\omega^2 + i\omega(\gamma_{k_F}^{c,ep} + \gamma_{k_F}^{c,ee})}. \label{eq:chi_intra}
\end{align}
The plasma frequency $\omega_{pl}$ is hereby given by the microscopic expression
\begin{align}
    \omega_{pl}^2 = \frac{2e^2}{3V\varepsilon_0 k_B T}\sum_{P,\mathbf{k}\in C^P}|\mathbf{v}_{k_{\perp}k_{\parallel}\phi}|^2f_{k_{\perp}k_{\parallel}}^{cP,eq}(1-f_{k_{\perp}k_{\parallel}}^{cP,eq})~. \label{eq:results_omega_pl_micro}
\end{align}
For the three fit ranges (i-iii), the evaluated plasma frequencies are listed in Tab.~\ref{tab:omega_plasma}. In comparison, the macroscopic definition of the plasma frequency \cite{bittencourt_fundamentals_2004} $\omega_{pl,cl}= \sqrt{e^2n_0/m\varepsilon_0} $, results in $\hbar\omega_{pl,cl} = 9.08~\text{eV}$.

\begin{table}[H]
    \caption{\label{tab:omega_plasma}%
    Plasma frequencies estimated by the fit ranges (i-iii).}
    \begin{ruledtabular}
        \begin{tabular}{cccc}
            Fit Range & (i) &(ii) & (iii)\\
            \hline
            $\hbar\omega_{pl}$ & $9.7 ~\text{eV}$ & $ 12.4 ~\text{eV}$ & $ 17.2 ~\text{eV}$
        \end{tabular}
    \end{ruledtabular}
\end{table}

\begin{figure}[ht!]
    \centering
    \includegraphics[width=\linewidth]{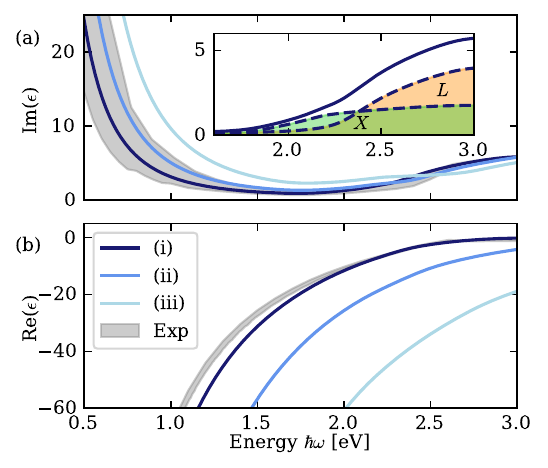}
    \caption{\justifying The (a) imaginary and (b) real part of the total optical permittivity at room temperature for three fit ranges of the anisotropic model (cf. Fig.~\ref{fig_dispersion}) is compared to compiled experimental data (grey bounded area) \cite{johnson_optical_1972,babar_optical_2015}. For the fit range (i), the inset shows the interband absorption edge decomposed in high symmetry point contributions.}
    \label{fig:spektrum}
\end{figure}

For interband excitations, the momentum dependence of the dephasing rate $\gamma_{P-k_{\parallel}}^{p,ee/ep}$ on $k_{\parallel}$ is maintained, since the occupation difference $f_{k_{\perp}k_{\parallel}}^{cP,eq} - f_{k_{\perp}k_{\parallel}}^{vP,eq}$ in Eq.~(\ref{eq:interband_susceptibility}) is finite for all momentum states above the Fermi level. \\
The total permittivity $\epsilon(\omega)$, cf. Eq.~(\ref{eq:chi_tot}), is shown  for the three fit ranges (i-iii) in Fig.~\ref{fig:spektrum}. The dipole matrix element $|\mathbf{d}|^2 = |\mathbf{d}^X|^2 = |\mathbf{d}^L|^2$, cf. Eq.~(\ref{eq:interband_susceptibility}), is hereby the only fit parameter, see Tab.~\ref{tab:fit_parameter_ranges}.
\begin{table}[H]
    \caption{\label{tab:fit_parameter_ranges}%
    Parameters of the linear spectrum in Fig.~\ref{fig:spektrum} yielded by the fit ranges (i-iii).}\begin{ruledtabular}
        \begin{tabular}{cccc}
            Fit Range & (i) &(ii) & (iii)\\
            \hline
            $|\mathbf{d}|^2$ & $0.015 \text{(eC nm)}^2$ & $0.036  \text{(eC nm)}^2$ & $0.112  \text{(eC nm)}^2$ \\
            $\epsilon_{\infty}$ & 7.3 & 5.7& 2.9 
        \end{tabular}
    \end{ruledtabular}
\end{table}

The background permittivity $\epsilon_{\infty}$, Tab.~\ref{tab:fit_parameter_ranges}, is determined from the difference $\epsilon_{\infty} = \epsilon_{\infty}^D - \chi_{\text{inter}}(0)$ of the Drude background permittivity $\epsilon_{\infty}^D$, taken from Ref.~\cite{vial_improved_2005}, and the low frequency limit of the interband susceptibility, Eq.~\eqref{eq:interband_susceptibility}.

In the near infrared regime, intraband processes dominate the optical response. For higher energies, starting at around 1.7 eV, interband transitions are excited and the interband susceptibility, Eq.~(\ref{eq:interband_susceptibility}), shapes the optical response. \\
The best overall agreement to experimental data \cite{johnson_optical_1972,magnozzi_plasmonics_2019} and to the plasma frequency is achieved by the fit range (i), taking into account the global properties of the electron dispersion rather than local features in the vicinity of high symmetry points as for the range (iii). Furthermore, the inset of Fig.~\ref{fig:spektrum} illustrates the link between the interband absorption edge and the different high symmetry point contributions to the interband susceptibility.
The discrepancies between the three fit ranges are particularly noticeable in the real part of the permittivity as a consequence of an overestimation of the plasma frequencies for the ranges (ii) and (iii), which can be traced back to a worse estimate of the DOS, cf. Fig.~\ref{fig_dos}.

\subsection{\label{sec:results:temperature_spectra}Linear Spectra as Function of Temperature}

\begin{figure*}[ht!]
    \centering
    \includegraphics[width=\linewidth]{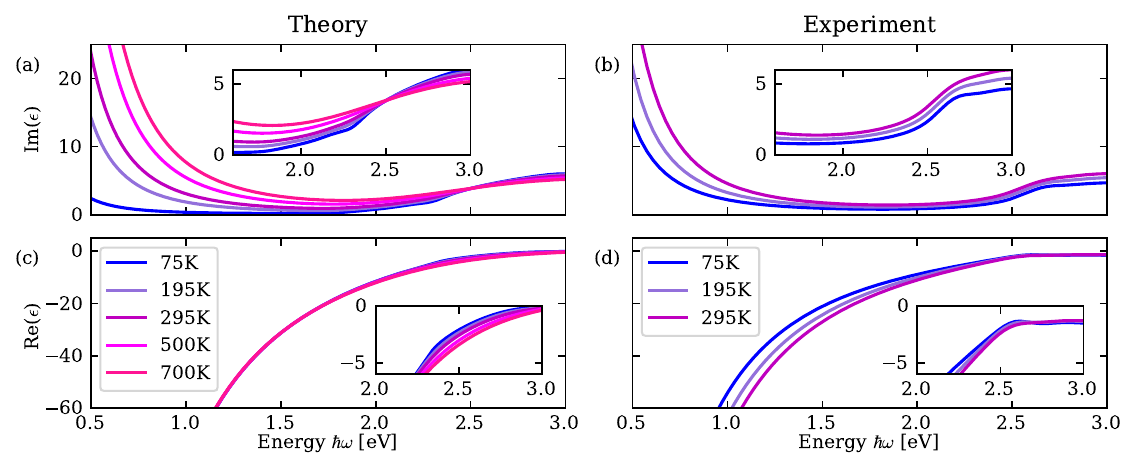}
    \caption{\justifying The (a) imaginary and (c) real part of the calculated total optical permittivity for temperatures in the range of 75~K to 700~K for fit range (i) (cf. Fig.~\ref{fig_dispersion}, Tab.~\ref{tab:anisotropic_fits_i}). In (b, d), this is compared to experimental results for temperatures in a range from 75~K to 295~K.}
    \label{fig:spektrum_temperature}
\end{figure*}

In the following, the fit range (i) is chosen as it provides the best agreement with experimental data at room temperature. The total permittivity $\epsilon(\omega)$ is calculated in Fig.~\ref{fig:spektrum_temperature}(a, c) for temperatures ranging from 75~K to 700~K and is compared to experimental results in Fig.~\ref{fig:spektrum_temperature}(b, d). We herefore conducted our own ellipsometry measurements on a polycrystalline gold film sample, cf.~App.~\ref{app:exp}.

\begin{figure}
    \centering
    \includegraphics[width=\linewidth]{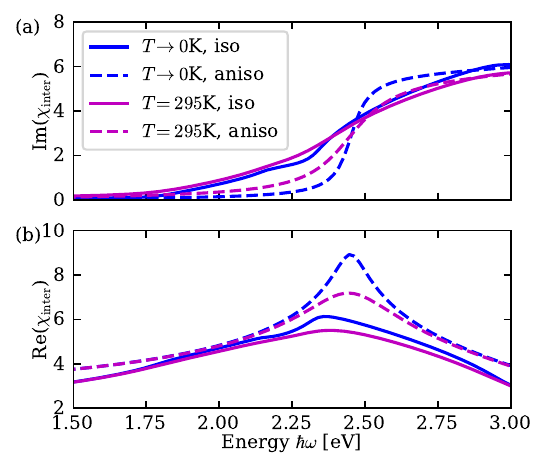}
    \caption{\justifying The (a) imaginary and (b) real part of the interband susceptibility are evaluated for the anisotropic (aniso, solid lines) and isotropic free electron dispersion model (iso, dashed lines) at room temperature and in the low temperature limit.}
    \label{fig:spectrum_limit}
\end{figure}

In the near infrared regime, our theory predicts a strong temperature sensitivity of the imaginary part of the intraband susceptibility, Eq.~\eqref{eq:chi_intra}, which vanishes for $T\rightarrow 0$. This reflects the temperature dependence of the relaxation rates $\gamma_{k_F}^{c}(T)$ in Fig.~\ref{fig:relaxation_rates}. In contrast, the real part, mostly determined by the temperature independent plasma frequency $\omega_{pl}$, does not change with temperature in Fig.~\ref{fig:spektrum_temperature}c.
The experimental results in Fig.~\ref{fig:spektrum_temperature}b show for the imaginary part a similar behavior as the theory prediction, whereby $\Im(\epsilon)$ approaches a finite value at low temperatures. This originates from additional losses caused by the polycrystalline structure and finite thickness of the sample \cite{yakubovsky_optical_2017, reddy_temperature-dependent_2016, olmon_optical_2012}. The real part increases in amplitude in the near infrared with temperature, which is in contrast to the theory prediction. In literature \cite{young_frequency_1969, reddy_temperature-dependent_2017}, this is attributed to temperature-dependent changes of the plasma frequency and effective mass not considered in our theory framework. 
\\
The calculated interband susceptibility on the other hand is less temperature dependent, in good agreement with the experimental results of Fig.~\ref{fig:spektrum_temperature}(b,d) and other studies \cite{zhang_dielectric_2018,reddy_temperature-dependent_2016,xu_role_2017,pells_optical_1969}. For the imaginary part, our theory shows an intersection point at which $\Im(\epsilon)$ is independent of $T$. This observation is also reported in the literature, mainly for silver, \cite{winsemius_temperature_1973, sundari_temperature_2013, ferrera_temperature-dependent_2019} but not observed in the experimental results in Fig.~\ref{fig:spektrum_temperature}b. At energies above $2.8$~eV, lower valence bands not considered in the two band model are activated by the optical excitation, which might explain the differences between theory and experiment at these energies. The weak dependence on temperature in the higher energy regime indicates that the broadening of the interband absorption edge cannot be exclusively attributed to the dephasing rate $\gamma_k^{p}$.
\\
The origin of the spectrally extended interband absorption edge, as already suggested in \cite{rosei_d_1973,guerrisi_splitting_1975}, is rather due to the anisotropic properties of the electronic band structure. This is illustrated in the inset of Fig.~\ref{fig:spektrum}a and in Fig.~\ref{fig:spectrum_limit}, where we compare the interband susceptibility of the anisotropic model in Eq.~(\ref{eq:interband_susceptibility}) with an analogous expression derived for the isotropic electron dispersion model, Eq.~(\ref{eq:dispersion:dispersion_iso}),
\begin{align*}
    &\chi_{\text{inter}}^{\text{iso}}(\omega) = \numberthis \label{eq:chi_inter_iso}\\
    & \frac{2}{3\varepsilon_0 \pi^2}\int_0^{q_D} dk k^2  \frac{|\mathbf{d}|^2(f_k^{c,eq} - f_k^{v,eq})(\varepsilon_k^{c} - \varepsilon_k^v)}{(\hbar\omega)^2 - (\varepsilon_k^c - \varepsilon_k^v)^2 + 2i\hbar\omega\hbar\gamma_k^p - (\hbar\gamma_k^p)^2}~.
\end{align*}
The dipole matrix element is chosen here to be identical to fit range (i), cf. Tab.~\ref{tab:fit_parameter_ranges}.\\
In Fig.~\ref{fig:spectrum_limit}, in the case of the anisotropic interband susceptibility (solid lines), the difference of the broadening of the absorption edge between room temperature and the low temperature limit $T \rightarrow 0$ is negligible. There, only a sharper crease caused by the different high symmetry point contributions is noticeable, cf. inset in Fig.~\ref{fig:spektrum}(a). In contrast, in the case of the isotropic interband susceptibility (dashed lines), the temperature is related to a significant broadening of the absorption edge. Especially in the low temperature limit, the spectrally extended absorption edge appearing in the experimental results in Fig.~\ref{fig:spektrum_temperature}b can not be reproduced.  This originates from the fact that in the isotropic approach mostly the finite electron-phonon dephasing rate contributes to the broadening, cf. Fig.~\ref{fig:dephasing_rates}(b), but less band structure related effects.
For lower temperatures, temperature-dependent many-body broadening is thus secondary in the origin of the spectrally extended absorption edge in the optical energy regime.

\section{Conclusion and Outlook}
We introduced metal Boltzmann-Bloch equations as a microscopic framework for the combined description of intra- and interband processes of optically induced electrons in noble metals. 
The MBBE are based on an anisotropic electronic dispersion model, suited to account for the geometrical features of energy surfaces of the partially filled conduction band. In combination with semi-analytical momentum-resolved temperature dependent expressions for the electron-electron and electron-phonon relaxation and dephasing rates, we discussed the linear intra- and interband optical response on the example of bulk gold and for different approaches on the anisotropic dispersion and temperatures.
A good agreement with experimental data of the temperature dependent dielectric function was hereby achieved by taking into account the global features of the band structure.
This substantiated the dominant role of electron-phonon processes in the total relaxation and dephasing rates.
In contrast to phenomenological Drude-Lorentz models, the MBBE approach allowed us to reduce the number of parameters required to fit experimental data significantly.\\
The developed framework motivates further studies on the interplay of intra- and interband processes for nonlinear excitations far from equilibrium and in plasmonic nanostructures.

\section*{Acknowledgments}
We acknowledge fruitful discussion with Lara Greten, Joris Sturm, Sabrina Meyer, Michiel Snoeken and Henry Mittenzwey (TU Berlin).
This work was supported by the German Science Foundation (DFG) under CRC/SFB 1636 – Project ID 510943930 - Project No. A08.
AK and JS acknowledge support from the Deutsche Forschungsgemeinschaft (DFG) through Project No. 527838492.
MR acknowledges support for the cryostat integration into the ellipsometer by the IBB Transfer BONUS programme No. 1657.

\appendix

\section{Phonon Weyl Operators}\label{app:weyl}
The equation of motion for the coherent phonon amplitude, resulting from the Hamiltonian  in Eq.~(\ref{eq:hamiltonian}) initially reads (without considering $H_{ep}^{W}$) 
\begin{align}
\partial_t \langle b_{\mathbf{q}}\rangle &=  -i\omega_{\mathbf{q}}\langle b_{\mathbf{q}}\rangle - \frac{i
}{\hbar}\sum_{\lambda\sigma\mathbf{kG}}g_{-\mathbf{qG}}^{\lambda}\langle a_{\mathbf{k-q+G}}^{\lambda\sigma\dagger}a_{\mathbf{k}}^{\lambda\sigma}\rangle~. \label{app:eq:eom_phonon_amplitude}
\end{align}
The right hand side of Eq.~(\ref{app:eq:eom_phonon_amplitude}) features electron-phonon coupling contributions without prior optical excitation and thus describes a non-equilibrium situation.\\
This can be addressed by a Weyl transformation of the coherent phonon amplitudes \cite{weber_optical_2007}
\begin{align}
    \Tilde{b}_{\mathbf{q}} = b_{\mathbf{q}} + c_{\mathbf{q}},
\end{align}
with the multiplicative offset operator 
\begin{align}
    c_{\mathbf{q}} = \frac{1}{\hbar\omega_{\mathbf{q}}}\sum_{\lambda\sigma\mathbf{kG}}g_{-\mathbf{qG}}^{\lambda}\langle a_{\mathbf{k-q+G}}^{\lambda\sigma\dagger}a_{\mathbf{k}}^{\lambda\sigma}\rangle^{eq}.
\end{align}
For readability, we relabel the Weyl operators as the original phonon operators ($\Tilde{b}_{\mathbf{q}}^{(\dagger)} \rightarrow b_{\mathbf{q}}^{(\dagger)}$). For the Hamiltonian in Eq.~(\ref{eq:hamiltonian}) follows
\begin{align*}
    H_{\text{W}}^{ep} = \sum_{\substack{\lambda\sigma\\\mathbf{qkG}}}g_{\mathbf{qG}}^{\lambda}\{ (b_{\mathbf{q}} + b_{-\mathbf{q}}^{\dagger})&\langle a^{\lambda\sigma\dagger}_{\mathbf{k+q+G}} a^{\lambda\sigma}_\mathbf{k}\rangle^{eq}\numberthis \label{app:weyl:hamiltonian} \\
    +& 2c_{\mathbf{q}}a_{\mathbf{k+q+G}}^{\lambda\sigma\dagger}a_{\mathbf{k}}^{\lambda\sigma}\}~.
\end{align*}

\section{Boltzmann-Bloch Scattering Rates}\label{app:scattering}
The scattering rates in Eq. (\ref{eq:theo_scattering_occupation_scattering}) and Eq. (\ref{eq:theo_scattering_transition_scattering}) read

\begin{widetext}
\begin{align*}
    T_{\mathbf{k}}^{p,ee} = &-\frac{\pi}{\hbar}\sum_{\mathbf{q'k'}\lambda} V_{\mathbf{k'-q'-k}}^{vc}V_{\mathbf{q'}}^{\lambda\lambda}\{(1-f_{\mathbf{k+q'+G}}^{\lambda})
    f_{\mathbf{k}}^{\lambda'}f_{\mathbf{k'-q'-G'}}^{\lambda} + f_{\mathbf{k+q'+G}}^{\lambda}(1-f_{\mathbf{k}}^{\lambda'})(1-f_{\mathbf{k'-q'-G'}}^{\lambda})\}\times\\
    &\hspace{5cm}\times\delta(\varepsilon_{\mathbf{k}}^{\lambda'} + \varepsilon_{\mathbf{k'}}^{\lambda} - \varepsilon_{\mathbf{k+q'+G}}^{\lambda} - \varepsilon_{\mathbf{k'-q'-G'}}^{\lambda'})p_{\mathbf{k'}}^{vc}\\
    & + \frac{\pi}{\hbar}\sum_{\mathbf{q'k'}\lambda}V_{\mathbf{k'-q'-k}}^{vc}V_{\mathbf{q'}}^{\lambda'\lambda}\{f_{\mathbf{k}}^{\lambda'}
    (1-f_{\mathbf{k+q'+G}}^{\lambda})(1-f_{\mathbf{k'}}^{\lambda'}) + f_{\mathbf{k+q'+G}}^{\lambda}f_{\mathbf{k'}}^{\lambda'}(1-f_{\mathbf{k}}^{\lambda'})\}\times\\
    &\hspace{5cm}\times\delta(\varepsilon_{\mathbf{k}}^{\lambda} + \varepsilon_{\mathbf{k'}}^{\lambda'} - \varepsilon_{\mathbf{k+q'+G}}^{\lambda'} - \varepsilon_{\mathbf{k'-q'-G'}}^{\lambda})p_{\mathbf{k'-q'-G'}}^{vc}~,  \numberthis \label{app:eq:scattering_T}
\end{align*}

\begin{align}
    \Gamma_{\mathbf{k}}^{\lambda,nl,ee} 
    = &\frac{\pi}{\hbar}\sum_{\mathbf{q'k'}}\{ (2V_{\mathbf{q'}}^{\lambda\lambda} - V_{\mathbf{k'-q'-k}}^{\lambda\lambda})V_{\mathbf{q'}}^{\lambda'\lambda}p_{\mathbf{k'-q'-G'}}^{\lambda\lambda'} p_{\mathbf{k'}}^{\lambda'\lambda}(f_{\mathbf{k+q'+G}}^{\lambda} - f_{\mathbf{k}}^{\lambda})\delta(\varepsilon_{\mathbf{k}}^{\lambda} + \varepsilon_{\mathbf{k'}}^{\lambda} - \varepsilon_{\mathbf{k+q'+G}}^{\lambda} - \varepsilon_{\mathbf{k'-q'-G'}}^{\lambda}) \nonumber \\
    & \qquad \quad + (2V_{\mathbf{q'}}^{\lambda'\lambda'} - V_{\mathbf{k'-q'-k}}^{\lambda'\lambda'})V_{\mathbf{q'}}^{\lambda'\lambda}p_{\mathbf{k}}^{\lambda\lambda'} p_{\mathbf{k+q'+G}}^{\lambda'\lambda}(f_{\mathbf{k'-q'-G'}}^{\lambda'} - f_{\mathbf{k'}}^{\lambda'})\delta(\varepsilon_{\mathbf{k}}^{\lambda'} + \varepsilon_{\mathbf{k'}}^{\lambda'} - \varepsilon_{\mathbf{k+q'+G}}^{\lambda'} - \varepsilon_{\mathbf{k'-q'-G'}}^{\lambda'}) \nonumber\\
    &\qquad \quad + 2V_{\mathbf{q'}}^{\lambda\lambda} V_{\mathbf{q'}}^{\lambda'\lambda}p_{\mathbf{k'-q'-G'}}^{\lambda'\lambda} p_{\mathbf{k'}}^{\lambda\lambda'}(f_{\mathbf{k+q'+G}}^{\lambda} - f_{\mathbf{k}}^{\lambda})\delta(\varepsilon_{\mathbf{k}}^{\lambda} + \varepsilon_{\mathbf{k'}}^{\lambda'} - \varepsilon_{\mathbf{k+q'+G}}^{\lambda} - \varepsilon_{\mathbf{k'-q'-G'}}^{\lambda'}) \nonumber \\
    &\qquad \quad + 2V_{\mathbf{q'}}^{\lambda\lambda} V_{\mathbf{q'}}^{\lambda'\lambda}p_{\mathbf{k+q'+G}}^{\lambda'\lambda} p_{\mathbf{k}}^{\lambda\lambda'}(f_{\mathbf{k'-q'-G'}}^{\lambda} - f_{\mathbf{k'}}^{\lambda})\delta(\varepsilon_{\mathbf{k}}^{\lambda'} + \varepsilon_{\mathbf{k'}}^{\lambda} - \varepsilon_{\mathbf{k+q'+G}}^{\lambda'} - \varepsilon_{\mathbf{k'-q'-G'}}^{\lambda}) \nonumber \\
    &\qquad \quad + (V_{\mathbf{q'}}^{\lambda\lambda} V_{\mathbf{k'-q'-k}}^{\lambda'\lambda}p_{\mathbf{k'-q'-G'}}^{\lambda'\lambda} p_{\mathbf{k}}^{\lambda\lambda'}(f_{\mathbf{k'}}^{\lambda} - f_{\mathbf{k+q'+G}}^{\lambda}) + V_{\mathbf{q'}}^{\lambda'\lambda}V_{\mathbf{k'-q'-k}}^{\lambda\lambda'}p_{\mathbf{k'}}^{\lambda'\lambda} p_{\mathbf{k}}^{\lambda\lambda'}(f_{\mathbf{k+q'+G}}^{\lambda} + f_{\mathbf{k'-q'-G'}}^{\lambda'}-1)) \times \nonumber \\
    & \qquad \qquad \qquad \qquad \qquad \qquad \qquad \times \delta(\varepsilon_{\mathbf{k}}^{\lambda'} + \varepsilon_{\mathbf{k'}}^{\lambda} - \varepsilon_{\mathbf{k+q'+G}}^{\lambda} - \varepsilon_{\mathbf{k'-q'-G'}}^{\lambda'}) \nonumber \\
    &\qquad \quad + (V_{\mathbf{q'}}^{\lambda\lambda} V_{\mathbf{k'-q'-k}}^{\lambda'\lambda}p_{\mathbf{k+q'+G}}^{\lambda'\lambda} p_{\mathbf{k'}}^{\lambda\lambda'}(f_{\mathbf{k}}^{\lambda} - f_{\mathbf{k'-q'-G'}}^{\lambda}) + V_{\mathbf{q'}}^{\lambda'\lambda}V_{\mathbf{k'-q'-k}}^{\lambda\lambda'}p_{\mathbf{k+q'+G}}^{\lambda'\lambda} p_{\mathbf{k'-q'-G'}}^{\lambda\lambda'}(1- f_{\mathbf{k}}^{\lambda} - f_{\mathbf{k'}}^{\lambda'})) \times \nonumber \\
    & \qquad \qquad \qquad \qquad \qquad \qquad \qquad \times \delta(\varepsilon_{\mathbf{k}}^{\lambda} + \varepsilon_{\mathbf{k'}}^{\lambda'} - \varepsilon_{\mathbf{k+q'+G}}^{\lambda'} - \varepsilon_{\mathbf{k'-q'-G'}}^{\lambda})\} + c.c.~,
\end{align}

\begin{align}
    &\Gamma_{\mathbf{k}}^{p,nl} = \nonumber\\
    \frac{\pi}{\hbar}\sum_{\mathbf{k'q'}\lambda} 
    &\{(2V_{\mathbf{q'}}^{\lambda\lambda} - V_{\mathbf{k'-q'-k}}^{\lambda\lambda})(V_{\mathbf{q'}}^{\lambda'\lambda'}p_{\mathbf{k'-q'-G'}}^{\lambda\lambda'}p_{\mathbf{k+q'+G}}^{\lambda\lambda'}p_{\mathbf{k'}}^{\lambda'\lambda} - V_{\mathbf{q'}}^{\lambda'\lambda}p_{\mathbf{k'}}^{\lambda\lambda'}p_{\mathbf{k}}^{\lambda\lambda'}p_{\mathbf{k'-q'-G'}}^{\lambda'\lambda})\delta(\varepsilon_{\mathbf{k}}^\lambda + \varepsilon_{\mathbf{k'}}^{\lambda} - \varepsilon_{\mathbf{k'-q'-G'}}^{\lambda}-  \varepsilon_{\mathbf{k+q'+G}}^{\lambda})\nonumber\\
    +&(2V_{\mathbf{q'}}^{\lambda'\lambda'} - V_{\mathbf{k'-q'-k}}^{\lambda'\lambda'})(V_{\mathbf{q'}}^{\lambda\lambda}p_{\mathbf{k'-q'-G'}}^{\lambda\lambda'}p_{\mathbf{k+q'+G}}^{\lambda\lambda'}p_{\mathbf{k'}}^{\lambda'\lambda} - V_{\mathbf{q'}}^{\lambda\lambda'}p_{\mathbf{k'}}^{\lambda\lambda'}p_{\mathbf{k}}^{\lambda\lambda'}p_{\mathbf{k'-q'-G'}}^{\lambda'\lambda})\delta(\varepsilon_{\mathbf{k'-q'-G'}}^{\lambda'}+ \varepsilon_{\mathbf{k+q'+G}}^{\lambda'}-\varepsilon_{\mathbf{k}}^{\lambda'} - \varepsilon_{\mathbf{k'}}^{\lambda'})\nonumber\\
    +&2V_{\mathbf{q'}}^{\lambda'\lambda}(V_{\mathbf{q'}}^{\lambda\lambda'}p_{\mathbf{k'-q'-G'}}^{\lambda'\lambda}p_{\mathbf{k+q'+G}}^{\lambda\lambda'}p_{\mathbf{k'}}^{\lambda\lambda'} - V_{\mathbf{q'}}^{\lambda\lambda}p_{\mathbf{k'}}^{\lambda'\lambda}p_{\mathbf{k}}^{\lambda\lambda'}p_{\mathbf{k'-q'-G'}}^{\lambda\lambda'})\delta(\varepsilon_{\mathbf{k'-q'-G'}}^{\lambda'}+\varepsilon_{\mathbf{k+q'+G}}^{\lambda}-\varepsilon_{\mathbf{k}}^{\lambda} - \varepsilon_{\mathbf{k'}}^{\lambda'})\nonumber\\
    +&2V_{\mathbf{q'}}^{\lambda\lambda'}(V_{\mathbf{q'}}^{\lambda'\lambda}p_{\mathbf{k'-q'-G'}}^{\lambda'\lambda}p_{\mathbf{k+q'+G}}^{\lambda\lambda'}p_{\mathbf{k'}}^{\lambda\lambda'} - V_{\mathbf{q'}}^{\lambda'\lambda'}p_{\mathbf{k'}}^{\lambda'\lambda}p_{\mathbf{k}}^{\lambda\lambda'}p_{\mathbf{k'-q'-G'}}^{\lambda\lambda'})\delta(\varepsilon_{\mathbf{k'-q'-G'}}^{\lambda}+\varepsilon_{\mathbf{k+q'+G}}^{\lambda'}-\varepsilon_{\mathbf{k}}^{\lambda'} - \varepsilon_{\mathbf{k'}}^{\lambda}) \label{app:eq:scattering_gamma_ee_p_nl}
    \}~,
\end{align}

\begin{align*}
    \Gamma_{\mathbf{k}}^{\lambda,nl,ep} = \frac{2\pi}{\hbar}\sum_{\mathbf{q'G\pm}}\pm|g_{\mathbf{q'G}}^{\lambda}g_{\mathbf{q'G}}^{\lambda'}|\text{Re}\{p_{\mathbf{k}}^{\lambda'\lambda}p_{\mathbf{k+q'}}^{\lambda\lambda'}\}
    \delta(\varepsilon_{\mathbf{k+q'+G}}^{\lambda} - \varepsilon_{\mathbf{k}}^{\lambda} \pm \hbar\omega_{\mathbf{q'}})~. \numberthis
\end{align*}

\end{widetext}

\section{Electron-Phonon and Electron-Electron Relaxation Rates}\label{app:rra_full}
The linearized electron-phonon occupation scattering contributions in Eq.~(\ref{eq:lin:occupation_rta}) can be approximated by an exponential decay, cf. Eq.~(\ref{eq:lin:occupation_gamma}), as shown in Ref.~\cite{czycholl_solid_2023}, by leveraging the properties
\begin{align}
      &f_{\mathbf{k'}}^{\lambda,eq}(1-f_{\mathbf{k}}^{\lambda,eq}) = e^{\frac{\varepsilon_{\mathbf{k}}^{\lambda}-\varepsilon_{\mathbf{k'}}^\lambda}{k_B T}}(1-f_{\mathbf{k'}}^{\lambda,eq})f_{\mathbf{k}}^{\lambda,eq},  \\
      &(f_{\mathbf{k}}^{\lambda,eq} - f_{\mathbf{k'}}^{\lambda,eq})n(\varepsilon_{\mathbf{k}}^\lambda -  \varepsilon_{\mathbf{k'}}^\lambda) = f_{\mathbf{k}}^{\lambda,eq}(1-f_{\mathbf{k'}}^{\lambda,eq}),
\end{align}
of the Fermi-Dirac distribution and making the Ansatz $f^{\lambda\sigma,1}_\mathbf{k} \approx -\frac{e}{k_B T} f^{\lambda,eq}_\mathbf{k}(1-f^{\lambda,eq}_\mathbf{k}) \mathbf{v}^\lambda_{\mathbf{k}}\cdot \int_{t_0}^t dt'  \mathbf{E}(t')$. This allows for an estimation of contributions involving the occupations $f_{\mathbf{k+q+G}}^{\lambda\sigma,1}$ in terms of $f_{\mathbf{k}}^{\lambda\sigma,1}$, the resulting relaxation rate is given by
\begin{align}
    \gamma_{\mathbf{k}}^{\lambda,ep} &= \frac{2\pi}{\hbar k_B T}\sum_{\mathbf{q}\pm}|g_{\mathbf{qG}}^{\lambda}|^2n(\hbar\omega_{\mathbf{q}})n(-\hbar\omega_{\mathbf{q}})\hbar\omega_{\mathbf{q}}\times\label{app:eq:rra_ep_general}\\
    &\times\frac{\{\mathbf{v}_{\mathbf{k+q+G}}^{\lambda}-\mathbf{v}_{\mathbf{k}}^\lambda\}\cdot\boldsymbol{\xi}}{\mathbf{v}_{\mathbf{k}}^\lambda\cdot\boldsymbol{\xi}}\delta(\varepsilon_{\lambda\mathbf{k+q+G}} - \varepsilon_{\lambda\mathbf{k}} \pm \hbar\omega_{\mathbf{q}})~.\nonumber
\end{align}
with $\boldsymbol{\xi}$ the polarization direction of the optical field.\\
In the case of the linearized electron-electron contributions in Eq.~(\ref{eq:lin:occupation_rta}) we follow the same procedure, and additionally leverage that the transferred energy is predominantly of the order of $k_B T$,  we can extract the electron-electron relaxation rates as 
\begin{align}
    \gamma_{\mathbf{k}}^{\lambda,ee} &= -\frac{2\pi}{(k_B T)^2\hbar}\sum_{\substack{\mathbf{qk'GG'}\\\sigma'\lambda'}}(2V_{\mathbf{q}}^{\lambda'\lambda} - V_{\mathbf{k-k'+q}}^{\lambda'\lambda}\delta_{\lambda\lambda'}^{\sigma\sigma'})V_{\mathbf{q}}^{\lambda\lambda'}\times\nonumber\\
    &\times n(\Delta E_{\mathbf{kqG}}^\lambda)n(-\Delta E_{\mathbf{kqG}}^\lambda)\Delta E_{\mathbf{kqG}}^{\lambda^2}f_{\mathbf{k'}}^{\lambda',eq}(1-f_{\mathbf{k'}}^{\lambda',eq})\times\nonumber\\
     &\times \frac{\{\mathbf{v}^\lambda_{\mathbf{k+q+G}} - \mathbf{v}^\lambda_{\mathbf{k}} + \mathbf{v}^{\lambda'}_{\mathbf{k'-q-G'}} - \mathbf{v}^{\lambda'}_{\mathbf{k'}}\}\cdot \boldsymbol{\xi} }{\mathbf{v}^\lambda_{\mathbf{k}}\cdot \boldsymbol{\xi}  }\times\nonumber\\
     &\times \delta(\Delta E_{\mathbf{kqG}}^\lambda + \varepsilon_{\mathbf{k'-q-G'}}^{\lambda'} -\varepsilon_{\mathbf{k'}}^{\lambda'}), \label{eq:app:rra_ee_general}
\end{align}
with the energy difference $\Delta E_{\mathbf{kqG}}^\lambda := \varepsilon_{\mathbf{k+q+G}}^\lambda - \varepsilon_{\mathbf{k}}^\lambda$.

\section{Evaluation of the isotropic Relaxation and Dephasing Rates}\label{app:rta}
\begin{figure}
    \centering
    \includegraphics[width=0.7\linewidth]{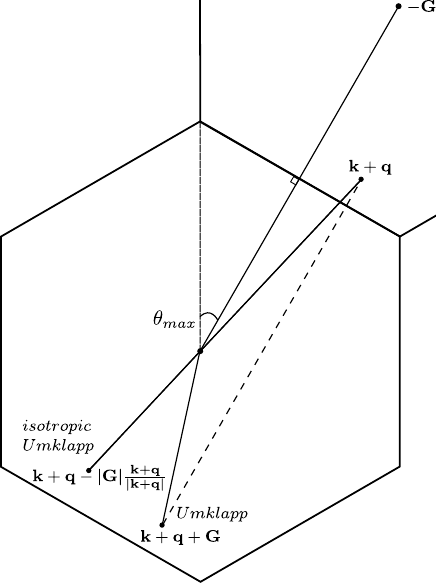}
    \caption{\justifying For a two dimensional Brillouin zone, an Umklapp process is compared to the corresponding isotropic Umklapp process. The angle between $\mathbf{k+q}$ and $\mathbf{G}$ is hereby upper bound the half angle $\theta_{max}$.}
    \label{app:fig:iso_umklapp}
\end{figure}

In the linear limit, we evaluate the analytical estimates of the dephasing and relaxation rates in Eq.~(\ref{eq:lin:occupation_gamma},\ref{eq:lin:transition_rta}) for the isotropic effective mass model, cp. Eq.~(\ref{eq:dispersion:dispersion_iso}), which we extend to include Umklapp processes as follows 
\begin{align}
    \varepsilon_{\mathbf{k+q+G}}^{\lambda} = \begin{cases}
        \frac{\hbar^2}{2m^{\lambda}}|\mathbf{k+q}|^2 + \epsilon^{\lambda}, &|\mathbf{k+q}| \leq q_D\\
        \frac{\hbar^2}{2m^{\lambda}}(|\mathbf{G}| - |\mathbf{k+q}|)^2 + \epsilon^{\lambda}, &|\mathbf{k+q}| > q_D
    \end{cases}. \label{app:eq:dispersion_iso}
\end{align}
The second case of Eq.~(\ref{app:eq:dispersion_iso}),  $|\mathbf{k+q}|>q_D$, hereby yields an approximation of $\varepsilon_{\mathbf{k+q+G}}^\lambda$, based on an isotropic Umklapp process  $\mathbf{k+q+G} \approx \mathbf{k+q} - |\mathbf{G}|\frac{\mathbf{k+q}}{|\mathbf{k+q}|}$, cp. Fig.~\ref{app:fig:iso_umklapp}. This approximation discards the geometrical features of the first Brillouin zone and corresponds to a small angle approximation $\cos\sphericalangle(\mathbf{k+q,G}) \approx 1$. This is justified as it holds $\cos\theta_{max}^X = 0.89$ and $\cos\theta_{max}^L = 0.77$ for the half angles  $\theta_{max}^X,\theta_{max}^L$ of the $X$ and $L$ cones, respectively. In Fig.~\ref{app:fig:iso_umklapp}, $\theta_{max}$ is illustrated for a two dimensional Brillouin zone.
For simplicity we estimate $|\mathbf{G}| \approx 2q_D$, which is twice the radius of the Brillouin zone.\\
For the estimation of the isotropic electron-phonon relaxation rates $\gamma_{k}^{\lambda,ep}$, cp. Eq.~(\ref{eq:lin:occupation_gamma}), we assume $\mathbf{k}$ to be parallel to the electrical field ($\mathbf{k}\parallel\boldsymbol{\xi}$).  We start with Eq.~(\ref{app:eq:rra_ep_general}) in spherical coordinates ($\sum_{\mathbf{q}} \rightarrow \frac{V}{(2\pi)^3}\int dq~d\phi ~d\cos\theta ~q^2$), with $\cos\theta = \cos\sphericalangle (\mathbf{k,q})$ and distinguish between normal and Umklapp contributions by considering either $\cos\theta \leq (q_D^2 - k^2 - q^2)/2kq$ or $\cos\theta > (q_D^2 - k^2 - q^2)/2kq$ respectively.    
In the case of normal processes the calculation is outlined in \cite{czycholl_solid_2023} and leads to the well known Bloch-Grüneisen formula \cite{czycholl_solid_2023,ashcroft_solid_1976} as in Eq.~(\ref{eq:lin:occupation_normal_rate_ep}). In the case of Umklapp processes, the isotropic Umklapp model in Eq.~(\ref{app:eq:dispersion_iso}) yields under energy conservation, ensured by the Delta function in Eq.~(\ref{app:eq:rra_ep_general}), approximately 
\begin{align}
    \frac{m^\lambda}{\hbar}\{\mathbf{v}_{\mathbf{k+q+G}}^\lambda - \mathbf{v}_{\mathbf{k}}^\lambda \}\cdot \boldsymbol{\xi} &= -\frac{4q_D^2 - q^2}{4q_D - 2k},\nonumber\\
    |\mathbf{q+G}|^2 &= k\frac{4q_D^2 - q^2}{2q_D - k} := (q+\hat{G})^2.
\end{align}
Here $\hat{G}$ is introduced to formally allow for the isotropic correspondence $|g_{\mathbf{qG}}^\lambda|^2 \rightarrow |g_{q\hat{G}}^\lambda|^2 $,
this leaves us with the expression in Eq.~(\ref{eq:lin:occupation_umklapp_rate_ep}). The lower bound hereby reflects, that (up to the phonon energy) the transferred phonon momentum needs to exceed $2(q_D - k)$ in order to ensure energy conservation.\\
The calculation of the normal and Umklapp isotropic electron-phonon dephasing rates in Eq.~(\ref{eq:lin:transition_normal_rate_ep}) and Eq.~(\ref{eq:lin:transition_umklapp_rate_ep}) follows the same procedure.\\
For the calculation of the isotropic electron-electron relaxation rates $\gamma_{k}^{\lambda,ee}$, cp. Eq.~(\ref{eq:lin:occupation_gamma}), we consider Eq.~(\ref{eq:app:rra_ee_general}) and neglect exchange contributions of the form $V_{\mathbf{k-k'+q}}$. We furthermore limit the calculations to the case $\lambda = c$ as the interest primarily lies on the conduction band occupations, we can additionally restrict $\lambda' = c$ as the Pauli blocking terms $f_{\mathbf{k'}}^{\lambda',eq}(1-f_{\mathbf{k'}}^{\lambda',eq})$ forbid contributions from the filled valence band.\\
Again for $\mathbf{k} \parallel \boldsymbol{\xi}$, we then distinguish the four cases of either $\mathbf{k+q}$ or $\mathbf{k'-q}$ lying either inside or outside of the Brillouin zone sphere by undergoing an Umklapp process. Within the isotropic dispersion model, the exclusively normal contribution vanishes, $\gamma_{k}^{c,ee,NN} = 0$, as in this case the total momentum is conserved
\begin{align}
    \mathbf{v}_{\mathbf{k+q}}^{c}-\mathbf{v}_{\mathbf{k}}^{c}+\mathbf{v}_{\mathbf{k'-q}}^c - \mathbf{v}_{\mathbf{k'}}^{c}= \mathbf{0}.
\end{align}
In spherical coordinates $\mathbf{k'}\rightarrow(k',\cos\theta_{k'},\phi_{k'})$,  $\mathbf{q}\rightarrow (q,\cos\theta_{q},\phi_{q})$, where $\theta_{k'},\theta_{q}$ are taken relative to $\boldsymbol{\xi}$, we adapt the energy difference $\Delta E_{\mathbf{kqG}}^c$ in Eq.~(\ref{eq:app:rra_ee_general}) to $\Delta E_{kq\cos\theta_q}^{c,N/U} $ to simplify the distinction between $\mathbf{k+q}$ normal and Umklapp processes. It hereby holds
\begin{align}
\Delta E_{kq\cos\theta_q}^{c,N} &= \frac{\hbar^2}{2m_c}(2kq\cos\theta_q + q^2), \\
\Delta E_{kq\cos\theta_q}^{c,U} &= \frac{\hbar^2}{2m_c}\{2kq\cos\theta_q  + q^2  \\
 & \qquad \qquad-  4q_D\sqrt{k^2 + q^2 + 2kq\cos\theta_q} +  4q_D^2\}. \nonumber
\end{align}
We then proceed to evaluate the $\mathbf{k',G',\sigma'}$ sums for the three remaining cases, which eventually leads to the contributions $\gamma_k^{c,ee,NU},\gamma_k^{c,ee,UN},\gamma_k^{c,ee,UU}$ in Eqs.(\ref{eq:lin:gamma_occupation_NU}-\ref{eq:lin:gamma_occupation_UU}) respectively.\\
The evaluation of the electron-electron dephasing rates $\gamma_{k}^{p,ee,NN}+\gamma_{k}^{p,ee,NU}$ and $\gamma_{k}^{p,ee,UN}+\gamma_{k}^{p,ee,UU}$ in Eqs.~(\ref{eq:lin:gamma_polarization_N}-\ref{eq:lin:gamma_polarization_U}) is done by following the same procedure.

\section{Susceptibility Tensors}\label{app:dyadic}

The linearized current density in frequency space as in Eq.~(\ref{eq:lin:j_explicit}) is in the general case not parallel to the electrical field as the dyadic velocity tensor $\mathbf{v}_{k_{\perp}k_{\parallel}\phi}^{cP}\otimes \mathbf{v}_{k_{\perp}k_{\parallel}\phi}^{cP}$ is involved in the expression. For a given high symmetry point $P$ the radial unit vector is defined as
\begin{align}
   \mathbf{e}_{\parallel}^{P} = {\overline{P\Gamma }}/\norm{\overline{P\Gamma }}.
\end{align}
We can then construct an orthonormal basis $\{\mathbf{e}_{\parallel}^{P}, \mathbf{e}_{\perp}^{P,1},\mathbf{e}_{\perp}^{P,2}\}$, such that $\mathbf{e}_{\perp}^{P,1}$ and $\mathbf{e}_{\perp}^{P,2}$ span the plane orthogonal to $\mathbf{e}_{\parallel}^{P} $. For an arbitrary definition of the angle $\phi \in [0,2\pi )$, the electron velocity in the conduction band is expressed as
\begin{align}
    \mathbf{v}_{k_{\parallel}k_{\perp}\phi}^{cP} = \frac{-\hbar}{m_{\parallel}^{cP}}k_{\parallel}\mathbf{e}_{\parallel}^P + \frac{\hbar}{m_{\perp}^{cP}}k_{\perp}(\cos\phi~\mathbf{e}_{\perp}^{P,1} + \sin\phi~\mathbf{e}_{\perp}^{P,2}).
\end{align}
It can be verified that the dyadic product of the electron velocity averaged over $\phi$ is of the form
\begin{align*}
   &\int_0^{2\pi}d\phi \mathbf{v}_{k_{\parallel}k_{\perp}\phi}^{cP} \otimes \mathbf{v}_{k_{\parallel}k_{\perp}\phi}^{cP} \\
   &\qquad = \hbar^2\pi\{ 2\bigg(\frac{k_{\parallel}}{m_{\parallel}^{cP}}\bigg)^2\mathbf{e}_{\parallel}^{P}\otimes \mathbf{e}_{\parallel}^{P} + \bigg(\frac{k_{\perp}}{m_{\perp}^{cP}}\bigg)^2(\mathbb{1} - \mathbf{e}_{\parallel}^{P}\otimes\mathbf{e}_{\parallel}^P )\}.\numberthis
\end{align*}
Lastly, when summed over all high symmetry points of one kind, i.e either over all $L$ or over all $X$ points, the following relations hold
\begin{align}
    \sum_{X}\int_0^{2\pi}d\phi \mathbf{v}_{k_{\parallel}k_{\perp}\phi}^{cX} \otimes \mathbf{v}_{k_{\parallel}k_{\perp}\phi}^{cX} = \frac{1}{3}\sum_{X}\int_0^{2\pi}d\phi |\mathbf{v}_{k_{\parallel}k_{\perp}\phi}^{cX}|^2, \label{eq:app:x_dyadic_identity}\\
     \sum_{L}\int_0^{2\pi}d\phi \mathbf{v}_{k_{\parallel}k_{\perp}\phi}^{cL} \otimes \mathbf{v}_{k_{\parallel}k_{\perp}\phi}^{cL} = \frac{1}{3}\sum_{L}\int_0^{2\pi}d\phi |\mathbf{v}_{k_{\parallel}k_{\perp}\phi}^{cL}|^2. \label{eq:app:l_dyadic_identity}
\end{align}
As the integrand $f_{k_{\perp}k_{\parallel}}^{cP,eq}(1-f_{k_{\perp}k_{\parallel}}^{cP,eq})/i\omega - (\gamma_{P - k_{\parallel}}^{c ,ep}+ \gamma_{P- k_{\parallel}}^{c ,ee})$ in Eq.~(\ref{eq:lin:j_explicit}) does not depend either on $\phi$ nor on the particular choice of the symmetry point, we can apply the identities in Eqs.~(\ref{eq:app:x_dyadic_identity},\ref{eq:app:l_dyadic_identity}), to obtain the scalar expression in Eq.~(\ref{eq:lin:chi_intra}) for the intraband susceptibility.\\
The same argument is made for the scalar expression of the interband susceptibility in Eq.~(\ref{eq:interband_susceptibility}) as we chose $\mathbf{d}^P \parallel \mathbf{e}^P_{\parallel}$.

\section{Experimental Section}
\label{app:exp}

\begin{figure}[t]
    \centering
    \includegraphics[width=1\linewidth]{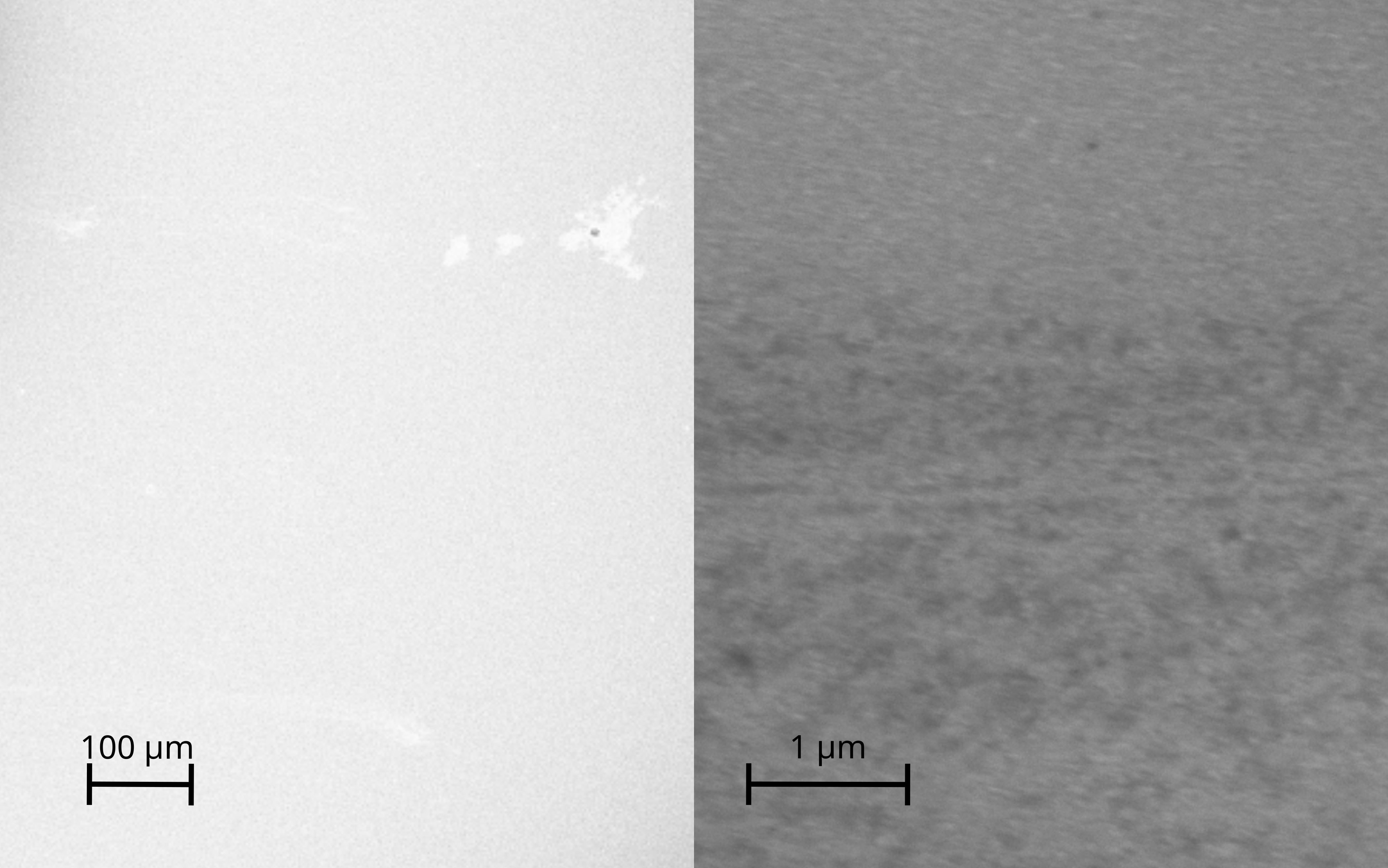}
    \caption{Scanning electron microscope image of the employed gold film. The film is polycrystalline with sub-$\mu$m domain sizes on the order of 100\,nm.}
    \label{fig:SEM}
\end{figure}

The optical response of gold was measured on a commercially available unprotected gold film (Thorlabs PF05-03-M03). Scanning electron microscopy images of the film are shown in Fig.~\ref{fig:SEM} showing crystallites on the 100\,nm length scale. 

Temperature-dependent spectroscopic ellipsometry spectra were acquired using a commercial SENTECH SENresearch SE850E ellipsometer, which has been equipped with a Leybold liquid Helium closed-cycle cryostat allowing to vary the temperature between $\approx15$\,K and $330$\,K. The sample was mounted on the cold head of the cryostat with a custom-made copper clamp. During the measurement, the sample was kept in high vacuum in order to avoid growing ice layers on the sample at low temperatures. Fused silica windows were mounted on rubber o-rings such that they do not alter the polarisation state of the probe light over the full measurement range $\hbar\omega=0.5-5$\,eV.

The ellipsometry spectra were obtained at the angle of incidence of the light of $\Phi=70^\circ$. The measurement software (SENTECH SpectraRay/3) yields directly the ellipsometric angles, $\Psi$ and $\Delta$, which describe the change of the light polarization upon reflection on the sample. Because of the large thickness of the gold layer, the ellipsometric angles can be directly inverted allowing to calculate the complex dielectric function  $\epsilon(\hbar\omega)=\epsilon_1+i\epsilon_2$ using the ellipsometric equation 
\begin{align}
\epsilon(\hbar\omega)=\sin^2\Phi+\sin^2\Phi\tan^2\Phi\left(\frac{1-\rho}{1+\rho}\right)^2
\end{align} 
with $\rho=\tan\Psi e^{i\Delta}$. 
The Kramers-Kronig consistency of the result for $\epsilon(\hbar\omega)$ was verified modeling the optical response using an optical multilayer model for the gold layer consisting of 8 oscillators in SpectraRay/3 and using a fused silica reference as substrate material in the model.

\bibliography{references}

\end{document}